\documentclass[a4paper,fleqn]{cas-dc}

\usepackage[numbers,sort&compress]{natbib}
\usepackage{booktabs}
\usepackage{graphicx} % Add the graphicx package for \resizebox
\usepackage{algorithm}
\usepackage{algorithmic}
\usepackage{hyperref}
\hypersetup{
    colorlinks=true,
    linkcolor=blue,
    filecolor=magenta,      
    urlcolor=cyan,
}

\def\tsc#1{\csdef{#1}{\textsc{\lowercase{#1}}\xspace}}
\tsc{WGM}
\tsc{QE}
\begin{document}
\let\WriteBookmarks\relax
\def\floatpagepagefraction{1}
\def\textpagefraction{.001}

\shorttitle{\emph{SW-ViT  for sequential multi-push USWE
}}  

% COMA-Net: Complementary Attention Network with Feature Shifting for Medical Image Segmentation

% \shorttitle{<short title of the paper for running head>} 

\shortauthors{\emph{AH Akash, MJ Alam and MK Hasan}}  

% \title[mode=title]{ Spatio-Temporal Vision Transformer Network  with Post Denoiser for Sequential Multi-Push Ultrasound Shear Wave Elastography: Simulation and Phantom Study}
\title[mode=title]{SW-ViT: A Spatio-Temporal Vision Transformer Network  with Post Denoiser for Sequential Multi-Push Ultrasound Shear Wave Elastography}

% \tnotemark[<tnote number>] 


%       facebook=<Facebook id>,

\author[1]{Ahsan Habib Akash}
\author[1]{MD Jahin Alam}

\author[1]{Md. Kamrul Hasan}%[type=editor,
    %   style=chinese,
    %   orcid=0000-0002-4816-2725]

\cormark[1]
% Footnote of the second author

% Email ID of the second author
% \ead{khasan@eee.buet.ac.bd}



\affiliation[1]{organization={Department of Electrical and Electronic Engineering, Bangladesh University of Engineering and Technology (BUET)},
            % addressline={West Palashi}, 
            city={Dhaka},
%          citysep={}, % Uncomment if no comma needed between city and postcode
            postcode={1205}, 
            country={Bangladesh}}

% \affiliation[2]{organization={Department of Electrical Engineering, West Virginia University},
%             % addressline={West Palashi}, 
%             city={Morgantown},
% %          citysep={}, % Uncomment if no comma needed between city and postcode
%             postcode={26505}, 
%             state={WV},
%             country={USA}}

\let\printorcid\relax
% \cortext[1]{Corresponding author}
\cortext[1]{Corresponding author: M. K. Hasan.\\
\-\ \-\ \-\ \-\ \-\ \-\ \-\ \-\ \-\ \emph{Email addresses}: khasan@eee.buet.ac.bd} 
% ## further enhancement Diagonastic quality
% ## Modular network
% ## Significance of the method
% ## Data description
\begin{abstract}
\textit{\textbf{Objective:}} Ultrasound Shear Wave Elastography (SWE) demonstrates great potential in assessing soft-tissue pathology by mapping tissue stiffness, which is linked to malignancy. Traditional SWE methods have shown promise in estimating tissue elasticity, yet their susceptibility to noise interference, reliance on limited training data, and inability to generate segmentation masks concurrently present notable challenges to accuracy and reliability.
\textbf{Approach:} In this paper, we propose \textbf{SW-ViT}, a novel two-step deep-learning framework for \textbf{S}hear \textbf{W}ave Elastography that integrates a CNN-Spatio-Temporal \textbf{Vi}sion \textbf{T}ransformer–based reconstruction network with an efficient Transformer–based post-denoising network. The reconstruction network uses a 3D ResNet encoder to extract hierarchical spatial and temporal features from sequential multi-push acoustic-radiation-force (ARF) imaging data, followed by multi-resolution spatio-temporal Transformer blocks-featuring both global and local attention for capturing temporal dependencies and converting 3D features into 2D representations. A squeeze-and-excitation, attention-guided decoder then reconstructs the 2D stiffness maps. To address limited data, we introduce a patch-based training modification that enables localized, region-wise training and reconstruction. In the second step, the post-denoising network refines these stiffness maps via a shared encoder and two decoders for foreground (inclusion) and background, whose outputs are fused into a denoised modulus map and a segmentation mask. A hybrid loss strategy combining regional supervision, smoothness, foreground–background fusion, and Intersection-over-Union (IoU) losses-ensures improvements in both estimation accuracy and segmentation quality. We validate the framework on both simulated and experimental multi-push SWE datasets. \textbf{Results:} Our method demonstrates robustness, achieving an average PSNR of $32.68,$dB, CNR of $46.78,$dB, and SSIM of $0.995$ on noisy simulated data, and PSNR of $21.11,$dB, CNR of $42.14,$dB, and SSIM of $0.936$ on an experimental CIRS049 phantom dataset. Segmentation performance is similarly strong, with IoU values of $0.949$ (simulation) and $0.738$ (phantom), and ASSD values of $0.184$ and $1.011$, respectively. Comparative analysis against existing deep-learning approaches confirms the superior performance of our method in noise mitigation and data efficiency. \textbf{ Significance:} This pipeline delivers robust, high-quality tissue elasticity estimates from noisy SWE data and holds clear promise for clinical application.

\end{abstract}.

 
\maketitle

% Main text
\section{Introduction}\label{introduction}

The mechanical characteristics of soft tissues (e.g. stiffness) are strongly correlated with pathological changes, making stiffness a valuable biomarker for diseases such as liver and kidney fibrosis, tendinopathies, and tumor development \cite{handorf2015tissue, martinez2021causal, piscaglia2016ultrasound, bob2017ultrasound, farrow2020novel, ferraioli2014shear, dirrichs2016shear}. Over the past decades, numerous techniques have been developed to quantify tissue elasticity for disease diagnosis and progression monitoring \cite{greenleaf2003selected}. Among these, elastography methods, especially strain elastography and shear-wave elastography (SWE), have demonstrated clinical utility \cite{shiina2015wfumb, barr2015wfumb}. Strain elastography, yielding only relative measurements, suffers from poor reproducibility, and is highly operator-dependent \cite{yoon2011interobserver}. In contrast, SWE provides absolute, operator-independent stiffness measurements with high reproducibility and reliability. SWE has been widely applied in clinical practice for assessing liver fibrosis, characterizing breast lesions, and detecting prostate cancer \cite{ferraioli2014shear, Youk, Woo}.

The SWE technique remotely induces tissue displacement using an acoustic radiation force (ARF) \cite{sarvazyan1998shear}. This displacement generates transient shear waves that propagate perpendicular to the ARF within the elastic medium. Under the assumptions of tissue incompressibility, isotropy, linearity, and elasticity, the local shear modulus, $\mu$, is related to the shear-wave speed, $v$, by  
\begin{equation}
\begin{aligned}
    \mu \;=\;\rho\,v^2
\end{aligned}
\end{equation}
where $\rho\approx1000\ \mathrm{kg\,m^{-3}}$ for soft tissue \cite{lai2009introduction}. Because the bulk modulus $K$ greatly exceeds $\mu$ $(K\gg\mu)$, the Young’s modulus, $E$, can be expressed as:  
\begin{equation}\label{sws}
E[Pa] = \mu \frac{3 K + 2 \mu}{K + \mu} \cong 3 \mu = 3 \rho \cdot v^2
\end{equation}

Various methods have been developed to determine shear-wave velocity using sequences of images that capture tissue displacements or velocity fields. These methods fall into two categories: time domain and frequency domain. The time-domain approach, often referred to as “Time of Flight” (ToF), involves measuring the time it takes for shear waves to travel a known distance. This is done by either tracking the wave peak via cross-correlation (CC) of two time-varying signals separated by a known distance \cite{tanter2008quantitative, song2012comb, bercoff2004supersonic, song2014fast}, or by performing a linear regression of the wave peaks in a 2D space–time plane \cite{palmeri2008quantifying, mclaughlin2006shear}. However, these approaches suffer in low signal-to-noise ratio (SNR) conditions, since noise, artifacts, and reflections at interfaces between tissues of differing stiffness can distort waveforms and bias velocity estimates \cite{6264136}. The frequency-domain approach estimates shear-wave velocity with phase-gradient or Fourier-transform techniques \cite{bernal2011material}. Phase-velocity dispersion due to material viscoelasticity and geometry has been characterized using methods in \cite{chen2009shearwave, kijanka2018robust, brum2014vivo}. Most implementations restrict analysis to a limited spatial window in both lateral and axial dimensions. Local Phase Velocity Imaging (LPVI) \cite{kijanka2018robust} further reconstructs velocity by exploiting sparsity in the local phase-velocity spectrum, yielding robust estimates from dominant wavenumbers. However, LPVI requires extensive tuning of imaging and filter parameters \cite{kijanka2018local, kijanka2019fastt}.

In recent years, deep learning methods have demonstrated robust noise resilience and high reconstruction quality in quantitative quasi-static ultrasound elastography, particularly strain elastography. These methods excel by learning to mitigate noise, as shown in several studies \cite{gao2020robust, liu2019deep, ma2021deep}. For example, Jush et al. \cite{jush2022deep} employed an encoder–decoder convolutional network to generate speed-of-sound maps directly from radio-frequency (RF) data, while Wu et al. \cite{wu2018direct} used a convolutional feature extractor and multilayer perceptron head to estimate pointwise displacement fields. Tripura et al. \cite{tripura2023wavelet} introduced a wavelet neural operator model to map tissue elastic properties. Furthermore, Tehrani et al. \cite{tehrani2022lateral} and Delaunay et al. \cite{delaunay2021unsupervised} applied unsupervised deep learning to quantify mechanical properties from strain-based RF data. These advances have significantly improved elasticity estimation in strain elastography. However, deep-learning applications in shear-wave elastography remain relatively underexplored.

Ahmed et al. \cite{ahmed2021dswe} demonstrated that deep learning models can generate both elasticity and segmentation maps from simulated tissue displacement data, and that models trained on simulations outperform traditional shear-wave speed (SWS) estimation methods on real-world phantom measurements \cite{kijanka2018local}. However, they noted that simulated noise does not accurately reflect the complex noise encountered during in vivo shear-wave acquisition. In contrast, Neidhardt et al. \cite{neidhardt2022ultrasound} trained and tested a 3D spatiotemporal convolutional neural network using experimental CIRS phantom data. Their model produces pixel-wise elasticity maps, even within the acoustic push region where conventional algorithms fail, yet remains sensitive to noise because each pixel’s estimate is computed independently. This independence limits coherence across neighboring pixels, and our experiments confirm that such localized estimation is more susceptible to noise, as adjacent estimates lack mutual regularization.

Several techniques have been developed to extend shear‐wave imaging over a larger region of interest (ROI) and to mitigate the blind spots that occur at each acoustic radiation force (ARF) application. Rather than using a single push beam, these methods employ multiple spatially separated ARF beams. Notable examples include comb-push ultrasound shear elastography (CUSE) \cite{song2012comb} and sequential multi-push shear-wave elastography (SWE) \cite{inoue2017development}. In CUSE, the transducer aperture is partitioned into subgroups that simultaneously transmit and track spatially displaced ARF beams. This overlap fills the blind zones of each individual push but introduces the challenge of managing constructive and destructive wave interference. By contrast, sequential multi-push SWE issues discrete ARF pushes at laterally offset positions, with tracking intervals between pushes which simplifies implementation by avoiding transducer subdivision but reduces the effective frame rate due to multiple tracking sequences. Our approach builds on the sequential multi-push paradigm by optimizing the lateral placement of each push so that overlapping ROIs receive only one ARF excitation at a time, thereby improving shear-wave data quality and enlarging the effective field of view (FOV).

\begin{figure*}[t]
  \centering
  \includegraphics[width=1\textwidth]{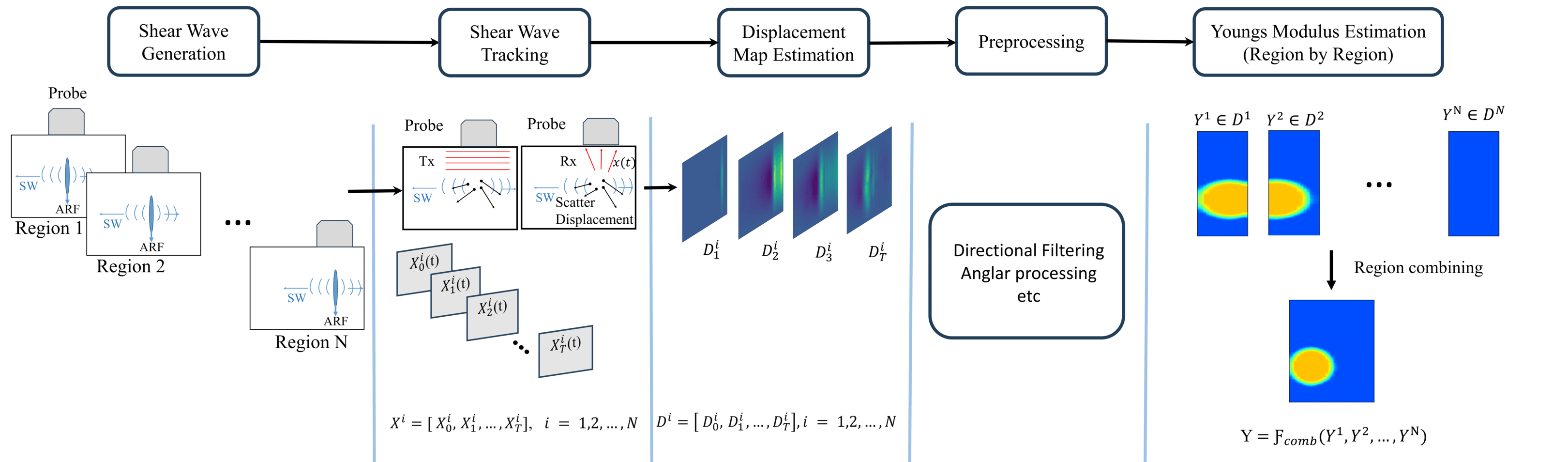}
  \caption{Working procedure of modulus estimation using multi-push shear wave elastography. The process begins with shear wave generation in different regions, followed by shear wave tracking using a probe. Displacement maps are then estimated and preprocessed through directional filtering and angular processing. Finally, Young's modulus is estimated for each region and combined to form the overall modulus map.}
  \label{methodology} 
\end{figure*}

In this paper, we present a novel two-step deep learning approach, termed `SW-ViT’, which integrates a CNN–spatio-temporal Vision Transformer reconstruction network with an efficient transformer-based post-denoiser for SWE. In the first step, a 3D ResNet encoder captures local spatial and temporal relationships by extracting hierarchical feature representations from each sub-region of the sequential multi-push ARF imaging sequence. These representations are fed into multiple spatio-temporal Transformer blocks at varying spatial and temporal resolutions, enabling the model to learn long-range dependencies across scales. We employ spatial-grid attention within these transformer blocks to refine temporal relation extraction at local spatial resolutions. A 3D convolutional layer then fuses the multi-scale feature space before passing it to the decoding stage. The decoder uses squeeze-and-excitation attention \cite{inproceedings} to reconstruct 2D stiffness maps from the learned motion features. To address limited medical data, we introduce a patch-based training strategy for the reconstruction network, which improves learning on individual sub-regions of each multi-push sequence. Finally, the reconstructed patches are spatially aggregated to form the full-ROI stiffness map.

In the second stage, inspired by Zamir et al. \cite{zamir2022restormer}, we implement an efficient transformer-based denoising network with a dual-decoder architecture for joint denoising and segmentation. A shared 2D encoder extracts feature maps from the initial ROI reconstruction, which are then fed into two parallel 2D decoders: one dedicated to foreground (inclusion) enhancement and the other to background refinement. Their outputs pass through two “Fusion” blocks: the first merges foreground and background features to produce a clean, high-quality modulus estimate; the second generates a binary segmentation mask of the inclusion region. We train this network using a multi-task loss that combines fusion consistency, structural similarity, smoothness regularization, and Intersection-over-Union (IoU) objectives. This compound training strategy enhances robustness and ensures precise separation of lesion and background.

We train and evaluate our method on data from sequential multi-push COMSOL simulations and CIRS phantom experiments. Our results demonstrate that SW-ViT outperforms existing reconstruction and denoising techniques, indicating its potential for applications that demand precise tissue characterization and reliable elasticity quantification. By leveraging advanced deep learning techniques, we aim to set a new standard in medical imaging and tissue analysis.

\section{Problem Formulation} \label{methods}

A shear wave is generated when acoustic radiation force (ARF) from an ultrasound transducer is applied to a soft-tissue medium. The wave propagates transverse to the push beam, inducing tissue displacement that is subsequently measured by the transducer (see Fig.~\ref{methodology}). In our study, we model scatterer motion in the axial–lateral plane, aligned with the tracking-beam propagation direction. For a single point scatterer located on the tracking line, the received RF signal is then described by the following model:
\begin{equation}\label{x_t}
x_0(t) = S(t - \tau_0) = A(t - \tau_0) \cos(\omega_c(t - \tau_0)) %, \;  {x}_0(t) \in \mathrm{X}_{\mathrm{0}}(t)
\end{equation}
Here, the subscripts index slow time (imaging frames) and the variable $t$ refers to fast time (RF sampling). The carrier angular frequency is denoted by $\omega_c$, and $\tau$ represents the two-way travel time between the transducer and the point scatterer. The function $S(\cdot)$ is the system impulse response, and $A(\cdot)$ is the real envelope of the received pulse. When a subsequent measurement is acquired after the scatterer has been displaced slightly by the propagating shear wave, the recorded RF signals are expressed as:

\begin{equation}
\begin{aligned}
  x_n(t) &= s(t - \tau_n) \\ 
  & = A(t - \tau_n) \cos(\omega_0(t - \tau_n)), \; n = 1, 2, \ldots, T_d
\end{aligned}
\end{equation}
where $\tau_n$ denotes the two-way travel time between the transducer and the scatterer after an axial displacement, relative to its initial travel time $\tau_0$. The scatterer’s displacement $d$ is directly proportional to the difference in these time delays, expressed as:
\begin{equation}\label{del_r}
\Delta r_n = c(\tau_{n-1} - \tau_n) = c\Delta\tau_n, \; n = 1, 2, \ldots, T_d 
\end{equation}
where $c$ represents the velocity of sound. Consequently, the absolute displacement of the single-point scatterer at time index $n$ is given by:
\begin{equation}\label{del_r}
r_{n} = \sum_{i=0}^{n} \Delta r_i 
\end{equation}
This single‐point tracking framework generalizes to a densely sampled two‐dimensional ROI by monitoring displacement fields across multiple scatterers, yielding:

\begin{equation}
U_n = \begin{bmatrix}
r_n(1,1) & r_n(1,2) & \cdots & r_n(1,a) \\
r_n(2,1) & r_n(2,2) & \cdots & r_n(2,a) \\
\vdots & \vdots & \ddots & \vdots \\
r_n(l,1) & r_n(m,2) & \cdots & r_n(l,a)
\end{bmatrix}
\end{equation}
Here, $U_n \in \mathbb{R}^{A \times L}$ denotes the 2D tissue displacement field at the $n$-th time frame, where $A$ and $L$ are the numbers of scatterers along the axial and lateral dimensions, respectively. By aggregating these displacement fields across all frames in an imaging sequence, we obtain a 3D volume $\mathbf{D}$, constructed by stacking the 2D particle velocity maps at successive time steps:

\begin{equation}\label{D}
\mathrm{\textbf{U}} = [U_1, U_2, \dots, U_{T_d}], \quad \mathrm{\textbf{U}} \in \mathbb{R}^{T_d \times A \times L } 
\end{equation}

\begin{figure*}[t]
  \centering
  \includegraphics[width=0.98\textwidth]{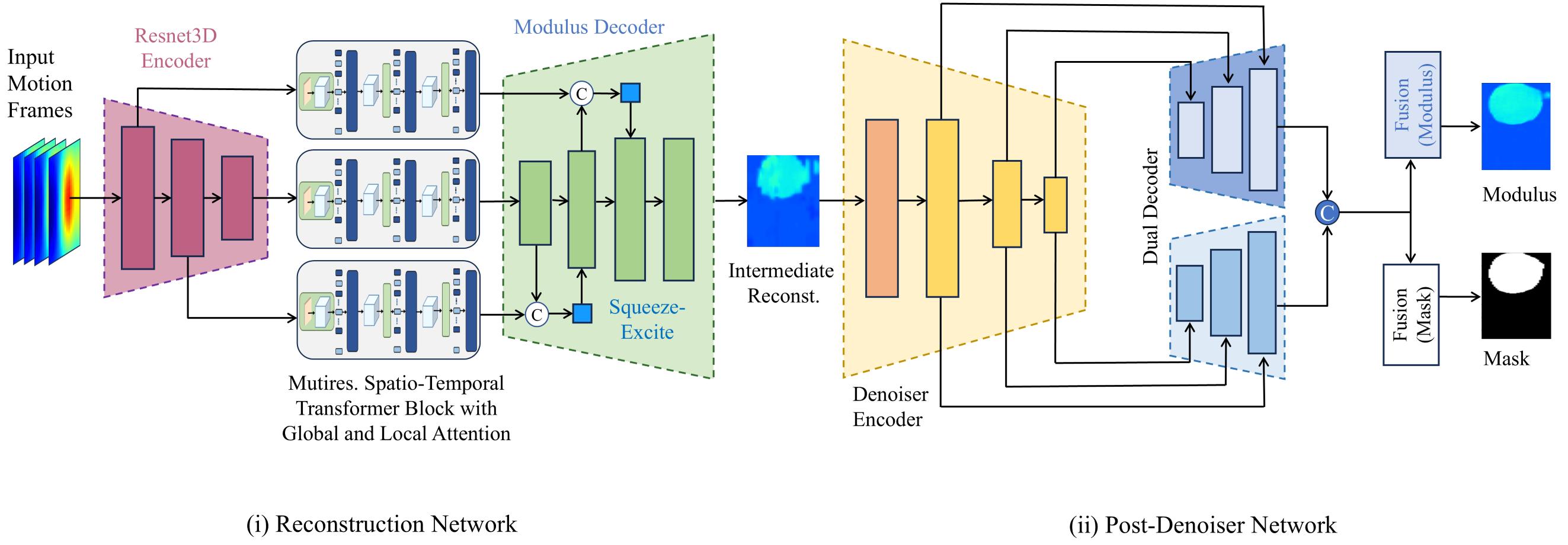}
  \caption{Proposed network Architecture: (i) the Reconstruction Network takes input motion frames processed by a Resnet3D encoder, followed by multi-resolution spatio-temporal ViT with global and local attention. The output is decoded by the Modulus Decoder, yielding an intermediate reconstruction. (ii) The Post-Denoiser Network further refines the intermediate reconstruction using a denoiser encoder, dual decoders, and fusion blocks, producing the final modulus and mask images.}
  \label{Complete_Network_block_diagram} 
\end{figure*}
Hence, the task of SWE imaging can be reduced to learning the transformations from displacement data, $\textbf{U}$, to a 2D Young's modulus (YM) map, $\mathrm{Y}'$. In our approach, this transformation is learned in two stages (see figure \ref{Complete_Network_block_diagram}). The first transfer function, $\mathcal{F}_{R}$ (i.e., reconstruction network), learns to map an intermediate YM directly from displacement data. The process for the reconstruction network can be defined mathematically as
\begin{equation}\label{Y}
\mathrm{Y}' = \mathcal{F}_{R}(\mathrm{\textbf{U}}; \theta_R), \mathrm{Y} \in \mathbb{R}^{ A \times L}
\end{equation}
where $\theta_R$ represents the learnable parameters of the reconstruction model and $\mathrm{Y}'$ is the intermediary reconstruction from a single region of the multi-push data. 

The quality and reliability of this reconstruction depend strongly on SNR of the propagating shear wave in the medium. Due to the high attenuation of shear waves, SNR decreases rapidly with distance from the push location. Increasing the ARF push strength to improve SNR is constrained by safety regulations, specifically the thermal index (TI) and mechanical index (MI), which limit maximum permissible ultrasound intensity \cite{nowicki2020safety}. Excessive intensity during SWE can increase the risk of tissue damage and compromise patient safety. To mitigate SNR loss while maintaining acceptable TI and MI, we employ a segmented imaging approach: the full ROI is divided into $N$ overlapping subregions, each acquired using a separate ARF push and imaging sequence (see Fig.~\ref{methodology}). Only a small subregion is insonified per push, ensuring acceptable SNR across each segment. Our reconstruction network processes these subregions individually. The complete ROI modulus image is then obtained by fusing the subregion estimates, $\mathrm{Y}^{i}$ and $\mathrm{M}^{i}$ $(i=1,2,\dots,N)$, via a tapered window function:

\begin{equation}
\mathrm{Y^i} = \mathcal{F}_{R} (\mathrm{\textbf{U}^i; \theta_R})
\end{equation}
\begin{equation}\label{Y_N}
\mathrm{Y'} = \mathcal{W} (\mathrm{Y^{1}}, \mathrm{Y^{2}}, \hdots, \mathrm{Y^{\textit{N}}}), \quad \mathrm{Y'} \in \mathbb{R}^{ A_{r} \times L_{r}}
\end{equation}
where $A_{r}\times L_{r}$ denotes the spatial dimensions of the full ROI. Subsequently, a denoising network refines the intermediate estimate to further enhance reconstruction quality. This refinement is expressed as
\begin{equation}\label{YM}
\left\{\mathrm{Y}, \mathrm{M}\right\} = \mathcal{F}_{D}(\mathrm{Y'}; \theta_D), \quad \left\{\mathrm{Y}, \mathrm{M}\right\} \in \mathbb{R}^{A_{r} \times L_{r}}
\end{equation}
where $\theta_{D}$ represents the learnable parameters of the denoiser model, $\mathrm{Y}$ is the final reconstructed YM map, and $M$ is the segmentation mask generated by the denoiser network.

\section{Methodology}
\subsection{Network Architecture}
Our deep-learning pipeline consists of two sequential networks: a reconstruction network and a post-denoising network. The reconstruction network takes the propagating shear wave displacement field as input and outputs an initial modulus estimate. The post-denoising network then refines this estimate to produce the final high-quality Young’s modulus map and corresponding segmentation mask. This two-stage architecture is illustrated in Fig.~\ref{Complete_Network_block_diagram}.

\subsection{Reconstruction Network}
The reconstruction network (Fig.~\ref{Network_block_diagram}) is designed to estimate Young’s modulus profiles from 3D volumetric shear wave displacement data. Its architecture is detailed below:

\begin{figure*}[!t]
  \centering
  \includegraphics[width=1\textwidth]{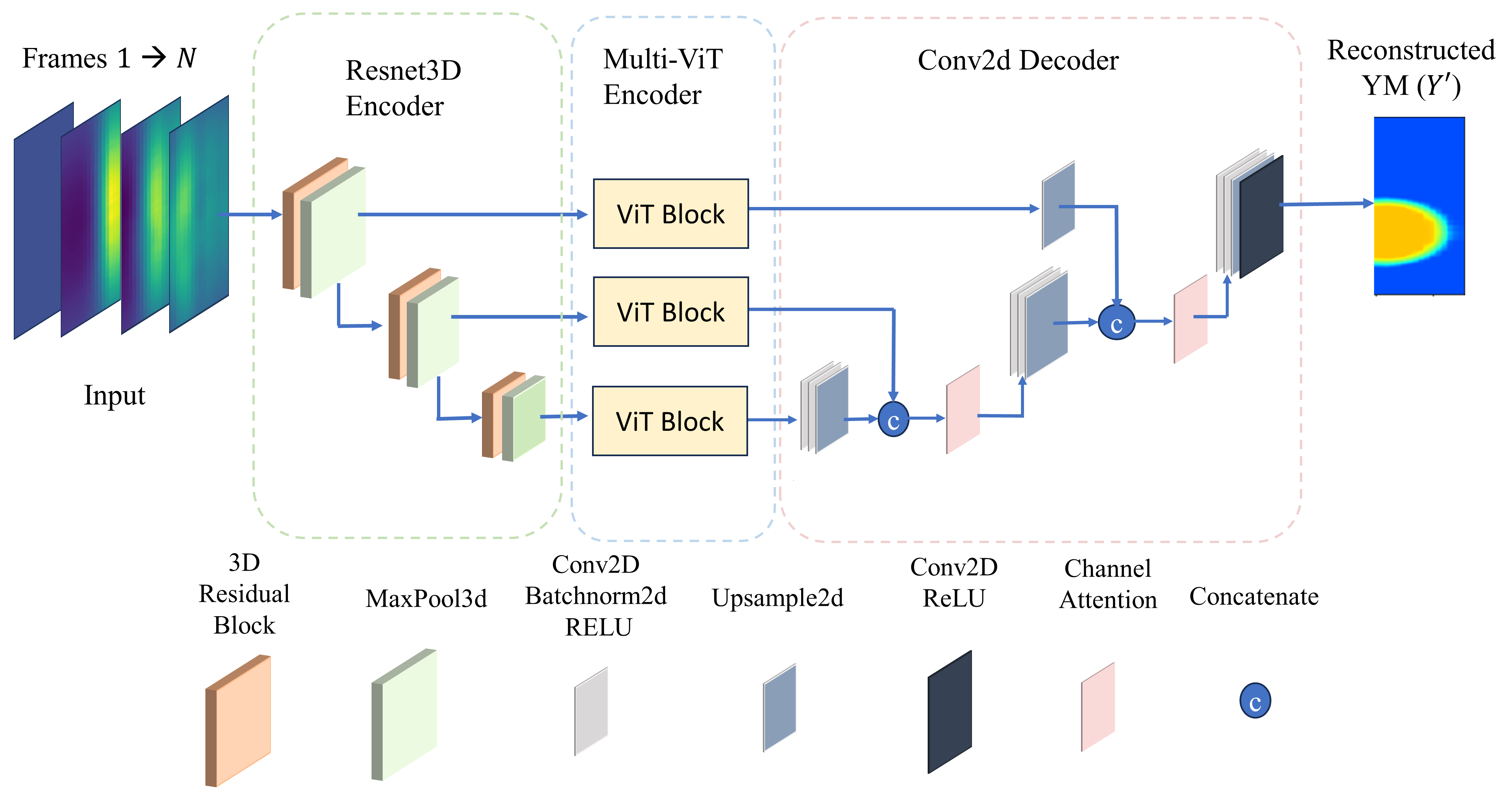}
  \caption{The proposed reconstruction network architecture. The process begins with input motion frames processed through a 3D residual block and MaxPool3d layer using a Resnet3D Encoder. The features are then passed through Multiple ViT (Vision Transformer) blocks. This is followed by a Conv2d Decoder with channel attention mechanisms.}
  \label{Network_block_diagram} 
\end{figure*}

\subsubsection{Residual Spatio-Temporal Enocoder} 

To robustly encode shear‐wave information from tissue particle displacements across multiple frames, we employ a 3D convolutional encoder composed of three residual stages based on 3D ResNets \cite{hara2017learning}. Each stage applies \(3\times3\times3\) convolutions followed by batch normalization, with 16, 32, and 64 kernels, respectively. After the first two stages, a \(2\times2\times2\) max‐pooling downsamples spatial and temporal dimensions, yielding feature maps \(\mathrm{I}_{0}\in\mathbb{R}^{C\times T/2\times A/2\times L/2}\) and \(\mathrm{I}_{1}\in\mathbb{R}^{2C\times T/4\times A/4\times L/4}\). In the final stage, we apply spatial-only \(2\times2\) pooling to preserve temporal resolution, producing \(\mathrm{I}_{2}\in\mathbb{R}^{4C\times T/4\times A/8\times L/8}\) with \(C=16\). This pooling strategy reduces dimensionality for richer feature extraction while retaining sufficient temporal information for the subsequent Spatio‐Temporal Vision Transformer block.

\begin{eqnarray}
\mathrm{I}_0 &=& \operatorname{P}_{2\times2\times2}(\mathcal{E}_{R0}(\mathrm{I}_{\mathrm{D,k}}; \mathrm{\theta}^e_0) + \mathrm{I}_{\mathrm{D}}), \notag \\
& & \mathrm{I}_0 \in \mathbb{R}^{C \times \frac{T}{2} \times \frac{A}{2} \times \frac{L}{2}} \label{eq:I_0}\\
\mathrm{I}_1 &=& \operatorname{P}_{2\times2\times2}(\mathcal{E}_{R1}(\mathrm{I}_0; \mathrm{\theta}^e_1) + \mathrm{I}_0), \notag \\
& & \mathrm{I}_1 \in \mathbb{R}^{2C \times \frac{T}{4} \times \frac{A}{4} \times \frac{L}{4}} \label{eq:I_1}\\
\mathrm{I}_2 &=& \mathrm{P}_{1\times2\times2}(\mathcal{E}_{R2}(\mathrm{I}_1; \mathrm{\theta}^e_2) + \mathrm{I}_1), \notag \\
& & \mathrm{I}_2 \in \mathbb{R}^{4C \times \frac{T}{4} \times \frac{A}{8} \times \frac{L}{8}} \label{eq:I_2}
\end{eqnarray}
Here, \(\mathcal{E}_{R0}\), \(\mathcal{E}_{R1}\), and \(\mathcal{E}_{R2}\) denote the three residual encoding blocks, \(\mathrm{P}\) denotes the pooling operation, and \(\theta^{e}_{0}\), \(\theta^{e}_{1}\), and \(\theta^{e}_{2}\) are the corresponding learnable parameters for each block.

\vspace{1\baselineskip}
\begin{figure*}[h]
\centering
\includegraphics[width=1\textwidth]{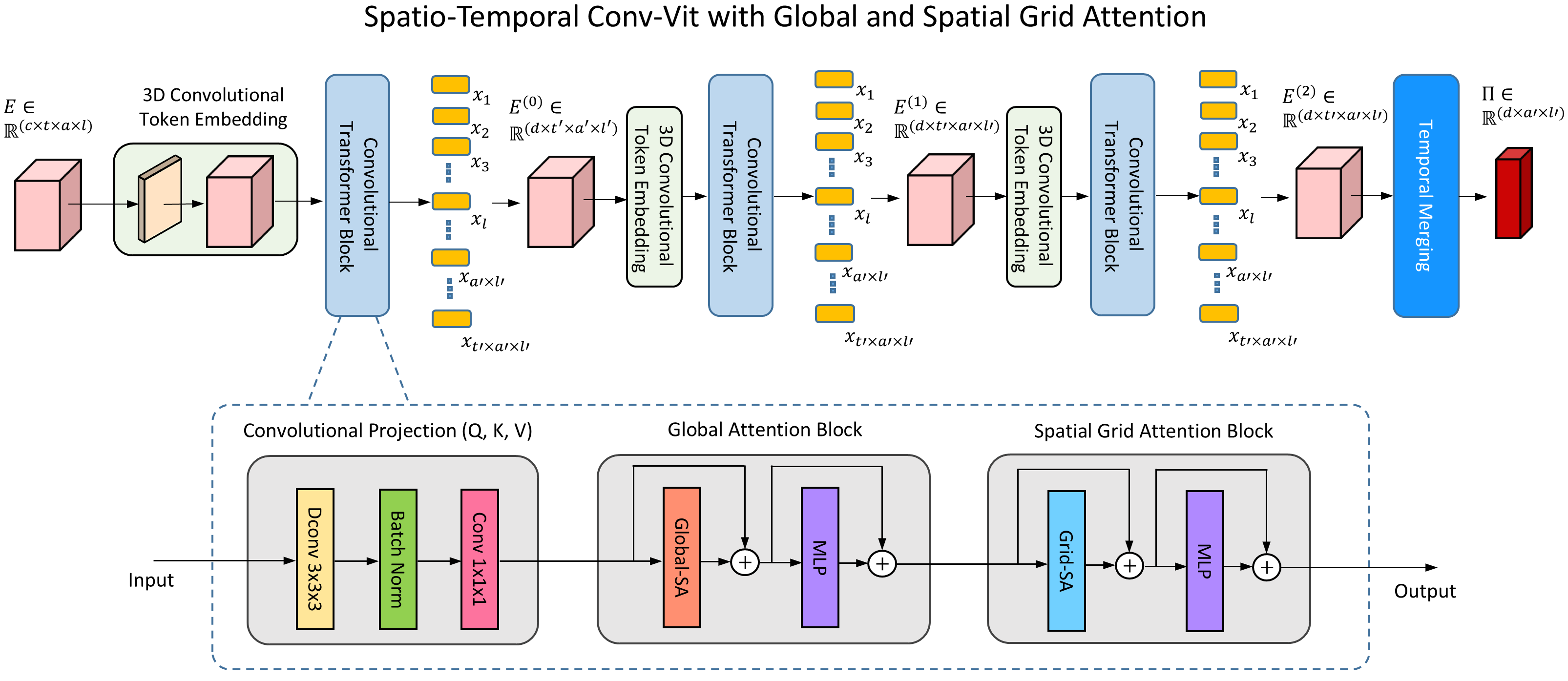}
\caption{Spatio-Temporal ViT with Global and Spatial Grid Attention. The architecture includes a Conv-Embedding Block that processes the input using 3D convolutional token embedding, batch normalization, and convolution layers. This is followed by multiple Convolutional Transformer Blocks. Each block consists of a combination of Global Attention and Spatial Grid Attention mechanisms to capture both global and localized spatio-temporal features. The output from these blocks is then fed into subsequent layers for further processing, leading to the final output.}
\label{Conv-Vit} 
\end{figure*}

\subsubsection{Multi-resolution Spatio-Temporal ViT} 

Capturing precise temporal and spatial correlations within shear‐wave displacement profiles is critical. To address this, we introduce a novel Spatio‐Temporal Convolutional Vision Transformer (ST‐ConvViT) block (Fig.~\ref{Conv-Vit}). By integrating this block at three resolution levels of the encoder’s feature maps, the model simultaneously captures fine‐grained spatial details and long‐range temporal dependencies. Each ST‐ConvViT block follows a consistent architecture optimized to model complex spatio‐temporal patterns, thereby enhancing the accuracy and robustness of shear wave speed estimation. The components of our Vision Transformer blocks are as follows:
\vspace{1\baselineskip}
\textbf{(i) 3D Convolutional Token Embedding:}
The 3D Convolution Token Embedding (3D‐CTE) block models local spatio‐temporal contexts effectively. At the $i$th stage, the residual encoder output $\mathrm{I}_i \in \mathbb{R}^{C_i \times T_i \times A_i \times L_i}$ is input to the 3D‐CTE, which applies a mapping \(\mathcal{F}_i(\cdot)\) to transform the channel dimension \(C_i\) into a new embedding:
\begin{equation}
    \mathrm{E}_{i} = \mathcal{F}_{i}(\mathrm{I}_{i}; \mathrm{\theta}^{f}_{i}),\quad \mathrm{E}_{i} \in \mathbb{R}^{C_{i} \times T_{i} \times A_{i} \times L_{i}}
\end{equation}
Here, \(\mathcal{F}_{i}(\cdot)\) denotes a 3D convolutional layer with a \(3\times3\times3\) kernel and stride \((s\times s\times s)\). By adjusting the stride and output channels of this convolution in the CTE block, both the token embedding dimension and the number of tokens at each stage can be controlled.
 
\vspace{1\baselineskip}
\textbf{(ii) Convolutional Projection:} Unlike the traditional Vision Transformer (ViT) \cite{dosovitskiy2021image}, which typically relies on linear projections, we propose  3D separable convolution to incorporate additional spatio-temporal context for each projection.  By leveraging 3D separable convolutions, we can efficiently capture the intricate dependencies across both spatial and temporal dimensions, enhancing the ability of the model to understand and process dynamic information. Figure \ref{Conv-Vit} shows our proposed Convolutional Projection Block (CPB). First, the tokens are reshaped to a 3D token map. Then the token map is passed to a depth-wise convolution (DConv) layer with kernel size  $\left(3 \times 3 \times 3\right)$ followed by a batch norm layer (BN) and finally a point-wise $\left(1 \times 1 \times 1 \right)
$ convolution layer (PWConv). The process is repeated three times to get the projection of Q, K, and V. This can be formulated as
\begin{align}
\mathrm{Q}_{n}^{h} &= {\Theta}_{Qn}^{h} \left(\mathrm{E}_{n}, \mathrm{\theta}_{qn}\right); \mathrm{Q}_{n}^{h}   \in \mathbb{R}^{d_{n} \times T_{n} \times A_{n} \times L_{n}}, \label{eq:Q_n}\\
\mathrm{K}_{n}^{h} &= {\Theta}_{Kn}^{h} \left(\mathrm{E}_{n}; \mathrm{\theta}_{kn}\right), \mathrm{K}_{n}^{h}  \in \mathbb{R}^{d_{n} \times T_{n} \times A_{n} \times L_{n}}, \label{eq:K_n}\\
\mathrm{V}_{n}^{h} &= {\Theta}_{Vn}^{h} \left(\mathrm{E}_{n}; \mathrm{\theta}_{vn}\right), \mathrm{V}_{n}^{h}  \in \mathbb{R}^{d_{n} \times T_{n} \times A_{n} \times L_{n}}. \label{eq:V_n}
\end{align}
Here, $\mathrm{Q}_{n}^{a}$ / $\mathrm{K}_{n}^{a}$ / $\mathrm{V}_{n}^{a} \in \mathbb{R}^{d_{n} \times T_{n} \times A_{n} \times L_{n}}$ are the query, key and value matrices at the given $n$-th layer of a transformer self-attention (SA) module, $h = [1, \ldots, \mathcal{H}]$ is an index over all attention heads. 

% \begin{equation}
% {Q',K',V'} = BN(\mathcal{F}_{DConv}(Reshape3D(\mathrm{I}_i), W_{dconv}))
% \end{equation}
% \begin{equation}
% {Q,K,V} = \mathcal{F}_{PWConv}((Q,K,V), W_{pwconv}))
% \end{equation}

 \textbf{(iii) Global and Local Spatio-temporal Self Attention:}  
Unlike convolutional models that emphasize local patterns, transformers excel at modeling global interactions via self‐attention. This global perspective is essential for accurate shear‐wave speed estimation, which requires correlating signals across the entire temporal domain to capture long‐range dependencies. To augment this capability, we integrate local spatial grid attention into the transformer architecture. This addition refines feature representations within localized regions while preserving the transformer’s global temporal context. By combining local and global attention mechanisms, our approach achieves more robust and precise shear‐wave speed estimation.

\textbf{(iii.a) Global Multi-Head Self Attention:} To leverage global dependencies, we reshape the initial matrices $\mathrm{Q}_{n}^{a}$ / $\mathrm{K}_{n}^{a}$ / $\mathrm{V}_{n}^{a} \in \mathbb{R}^{d_{n} \times T_{n} \times A_{n} \times L_{n}}$ into $\mathrm{Q}_{n}^{a}$ / $\mathrm{K}_{n}^{a}$ / $\mathrm{V}_{n}^{a} \in \mathbb{R}^{ \mathcal{V}_{n}  \times d_{n}}$, $ \mathcal{V}_{n} = (T_{n} \cdot A_{n} \cdot L_{n})$  Subsequently, these reshaped matrices undergo multi-head self-attention (MSA) block. The resulting vanilla self-attention map for each head is denoted as $\boldsymbol{\alpha}_{n}^{a} \in \mathbb{R}^{ \mathcal{V}_{n} \times \mathcal{V}_{n} }$, represents the "semantic similarity" between $\mathrm{Q}_{n}^{a}$ and  $\mathrm{K}_{n}^{a}$ in global context:
\begin{align}
\boldsymbol{\alpha}_{n}^{a} = \text{softmax}\left(\frac{\mathrm{Q}_{n}^{a^\top}}{\sqrt{D_k}} \cdot \mathrm{K}_{n}^{a}\right), \quad
\boldsymbol{\alpha}_{n}^{a} \in \mathbb{R}^{\mathcal{V}_{n} \times \mathcal{V}_{n} } 
\label{eq:alpha}
\end{align}
where $\text{softmax}$ denotes the softmax activation function, $\alpha^{a}_{n}$ is the attention map for the $a$-th head at position $n$. Next, we compute the weighted sum of the value vectors using the $\alpha$ coefficients:
\begin{equation}
\mathrm{s}_{n}^{a} = \boldsymbol{\alpha}_{n}^{a} \cdot \mathrm{V}_{n}^{a}, \quad  \mathrm{s}_{n}^{a}  \in \mathbb{R}^{ \mathcal{V}_{n} \times d_{n}}
\label{eq:sn}
\end{equation}
Finally, the concatenation of these vectors from all heads are passed through a multi-layer perceptron (MLP), incorporating residual connections from the previous block after each operation:
\begin{align}
\mathrm{S}_{n} &= \left[
\begin{array}{c}
\mathrm{s}_{n}^{1} \\ 
\vdots \\ 
\mathrm{s}_{n}^{(\mathcal{A})}
\end{array}
\right]  \\
\mathrm{S}_n^o &= \operatorname{MLP}\left(\operatorname{LN}\left(\mathrm{S}_{n}\right)\right) + \mathrm{S}_{n}
\end{align}
where, $\mathrm{S}_n^o$ is the final output of the global attention at $n$-th position, $\operatorname{LN}$ denotes the layer normalization operation.

\textbf{(iii.b) Local Spatial Grid Attention:} The model focuses on fine-grained feature details within local regions after capturing global dependencies. A local spatial grid attention mechanism is proposed for this purpose. Here, attention is applied specifically to a spatial grid, with the temporal dimension left unchanged since shear wave speed information is primarily embedded within the temporal information at a given spatial location.

To compute the local spatial attention, we reshape the initial matrices $\mathrm{Q}_{n}^{a}$ / $\mathrm{K}_{n}^{a}$ / $\mathrm{V}_{n}^{a} \in \mathbb{R}^{d_{n} \times T_{n} \times A_{n} \times L_{n}}$ into 
$\mathrm{Q}_{n}^{a}$ / $\mathrm{K}_{n}^{a}$ / $\mathrm{V}_{n}^{a} \in \mathbb{R}^{\frac{A_{n}}{H_{n}} \times \frac{L_{n}}{W_{n}} \times \mathcal{V}_{n}  \times d_{n}}$, $ \mathcal{V}_{n} = (T_{n} \cdot W_{n} \cdot H_{n})$.  $W_{n}$  and  $H_{n}$ are the axial and lateral window sizes respectively. Subsequently, these reshaped matrices undergo multi-head self-attention (MSA). Unlike the global attention, this resulting self-attention map for each head ($\boldsymbol{\alpha}_{n}^{a} \in \mathbb{R}^{ \mathcal{V}_{n} \times \mathcal{V}_{n}}$), represents the "semantic similarity" between $\mathrm{Q}_{n}^{a}$ and  $\mathrm{K}_{n}^{a}$ in a local context. The attention maps are generated similarly as shown in equation (\ref{eq:alpha}).

\textbf{(iv) Temporal Merging Block:}
After the STVT block, the feature tensor has a 3D spatio-temporal structure. Let $\text{X}_n \in \mathbb{R}^{d_{n} \times T_n \times A_n \times L_n}$
be the output of the STVT block. In order to prepare this tensor for the 2D decoder block, we collapse the temporal dimension using a 3D convolutional operation as follows. First, we swap the temporal and feature axes so that the temporal dimension becomes the channel dimension for the convolution. Define:
\begin{align}
\text{X}_n' = \operatorname{permute}(\text{X}_n) \in \mathbb{R}^{T_n \times d_{n} \times A_n \times L_n}
\end{align}
This rearrangement ensures that the temporal resolution \(T_n\) acts as the input channel for the subsequent convolution. Next, we apply a 3D convolutional layer with a kernel that covers the entire temporal depth. Let \(\mathcal{F}_{tm}(\cdot; \theta_{tm})\) denote the 3D convolution with kernel size \((T_n, 1, 1)\) and with output channels set to 1. The operation is given by:
\begin{align}
\pi_n = \mathcal{F}_{tm}\left(\text{X}_n'; \theta_{tm}\right) \in \mathbb{R}^{1 \times d_{n} \times A_n \times L_n}
\end{align}
In this convolution, the kernel spans the full temporal axis (i.e. \(T_n\)), effectively aggregating temporal information into a single channel while preserving the spatial dimensions. Finally, we remove the singleton channel resulting from the temporal merging to yield a 2D feature map:
\begin{align}
\Pi_n = \operatorname{Squeeze}\left(\pi_n\right) \in \mathbb{R}^{d_{n} \times A_n \times L_n }
\end{align}
The final 2D feature map \(\Pi_{n}\) now contains the spatially resolved features with the temporal information fully merged. It is passed as input to the attention guided decoder block.  
% \begin{align}
% \boldsymbol{\alpha}_{n}^{a} =\text{softmax}\left(\frac{\mathrm{Q}_{n}^{a^\top}}{\sqrt{D_k}} \cdot \mathrm{K}_{n}^{a}\right), \quad
% \boldsymbol{\alpha}_{n}^{a} \in \mathbb{R}^{\mathcal{V}_{n} \times \mathcal{V}_{n}} 
% \label{eq:alpha}
% \end{align}
% where, $V_{n} = (T_{n} \cdot W_{n} \cdot H_{n})$ is the length of the attention map for the local spatio-temporal context. 

\subsubsection{Attention Guided Decoder Block} The decoder module is structured to produce the primary elasticity estimation from the encoded features. It includes three stages of convolutional blocks, each accompanied by an upsampling step to reconstruct the modulus estimation to the original spatial dimensions. After each convolutional operation, a \(2 \times 2\) upsampling is performed to restore the spatial resolution.

The STViT block outputs $(\Pi_{(2-i)} \in \mathbb{R}^{2^{2-i}C \times \frac{A}{2^{2-i}} \times \frac{L}{2^{2-i}}}, \, i=0,1,2)$ are concatenated with the features from the previous decoder level $(\mathrm{D}_{(i-1)})$. This combined output is then processed through a Conv2D $\rightarrow$ BatchNorm $\rightarrow$ ReLU $(\mathcal{D}_{i})$ and then spatially upsampled (US). The upsampled features are further concatenated with the outputs from the next level of the STViT block $(\Pi_{1-i} \in \mathbb{R}^{B \times 2^{1-i}C \times \frac{A}{2^{1-i}} \times \frac{L}{2^{1-i}}})$ and passed through SE-attention $(\mathcal{A}_{i}(\cdot))$ blocks, which assign relevant weights to the most notable channels. This process can be mathematically described as:

\begin{equation}
    \mathrm{D}_{i} = \operatorname{US}_{2 \times 2}\left( \mathcal{D}_{i}\left(\mathrm{D}_{i-1}; {\theta}^{d}_{i}\right)\right), \;\; i=0,1,2
\end{equation}
\begin{equation}
  \mathrm{D}_i{ }^{\prime}=\left[\begin{array}{c}
\mathrm{D}_{i} \\
\Pi_{1-i}
\end{array}\right], \;\; i=0,1
\end{equation}
\begin{equation}
  \mathrm{D}^{SE}_{i} = \mathcal{A}_{i}\left(\mathrm{D'}_{i}; W_{i}^{a}\right), \;\; i=0,1
\end{equation}
Spatial upsampling is performed three times (\(i=0,1,2\)), whereas the other operations are executed twice (\(i=0,1\)). Consequently, \(\mathrm{D}_{2} \in \mathbb{R}^{C \times A \times L}\) is obtained. Finally, the output \(\text{Y}_{k}'\) is achieved as the primary modulus reconstruction:

\begin{equation}
    \text{Y}_{k}' = \mathcal{F}_{conv, ReLU}\left(\mathrm{D}_2; {\theta}_{conv}\right), \;\; \text{Y}_{k}' \in \mathbb{R}^{1 \times A \times L}
\end{equation}
Training the primary reconstruction network to generate \(\text{Y}_{k}'\) is feasible with a substantial SWE data pool. Optimal training of models relies on having ample data to accurately map tissue elasticity from displacement data. However, when data is limited, especially from tissue-mimicking phantoms or real-world \emph{in vivo} settings, acquiring large quantities of high-quality data can be challenging. This limitation can affect model performance and hinder optimization. To address this issue, we propose a patch-based training strategy, which is detailed in the following section.

\subsection{Patch-based Training} \label{Patch}

A reconstruction network requires large numbers of samples to learn the mapping from 3D motion frames to 2D stiffness maps. In practice, however, clinical SWE datasets are limited. Pixel-wise reconstruction using each spatial location in the motion tensor \(\mathbb{R}^{T\times A\times L}\) can be performed to predict its central modulus value, as in Neidhardt et al. \cite{neidhardt2022ultrasound}. However, this approach ignores neighborhood context and is highly sensitive to noise. Instead, we propose reconstructing local spatial patches from corresponding motion patches. By aggregating information over a patch, the model averages out uncorrelated noise at individual pixels, reducing variance and improving robustness.

To mitigate data scarcity in training the reconstruction network, we adapt its architecture to process smaller motion patches \(I_{(Dp,k)}\in\mathbb{R}^{1\times T\times A_p\times L_p}\) with \(A_p<A\) and \(L_p<L\), rather than the full motion volume. Although patching reduces the network’s receptive field, we compensate by training it to output correspondingly smaller stiffness maps \(Y'_p\in\mathbb{R}^{1\times\lceil A_p/3\rceil\times\lceil L_p/2\rceil-1}\), and by adjusting convolutional padding and using fractional upsampling within the encoder, STViT, and decoder. The feature sizes for such patch-based training are shown in Table \ref{Patch_based_train}. This patch‐based approach increases the number of effective training samples and enhances the network's robustness and generalization when mapping noisy motion data to precise modulus estimates.

% Suppose that the reconstructions within the FOV are taken with no spatial overlapping. In that case, the training data will increase six-fold within the patch alone, as $\textbf{Y}'_p$ covers almost \((1/3)\)-th of the axial and \((1/2)\)-th of the lateral dimension of the input patch. 

\begin{table}[h]
    \centering
    \caption{Feature sizes during patch-based training}
    \label{Patch_based_train}
    \begin{tabular}{cc}
    \hline
        \textbf{Feature} & \textbf{Size} \\  \hline
         $\mathrm{I}_{\mathrm{Dp,k}}$ & ${B \times 1 \times T \times A_p \times L_p}$ \\ [1mm]
         $\mathrm{I}_{\mathrm{0}}$ & ${B \times C \times \frac{T}{2} \times \left\lceil \frac{A_p}{3} \right\rceil \times \left\lceil \frac{L_p}{2} \right\rceil}$ \\ [1mm]
         $\mathrm{I}_{\mathrm{1}}$ & ${B \times 2C \times \frac{T}{4} \times \left\lceil \frac{A_p}{9} \right\rceil \times \left\lceil \frac{L_p}{4} \right\rceil}$ \\ [1mm]
        $\mathrm{I}_{\mathrm{2}}$ & ${B \times 4C \times \frac{T}{4} \times \left\lceil \frac{A_p}{9} \right\rceil \times \left\lceil \frac{L_p}{8} \right\rceil}$ \\ [1mm]
        % $\mathrm{P}_{\mathrm{0}}$ & ${B \times C \times \left\lceil \frac{A_p}{3} \right\rceil \times \left\lceil \frac{L_p}{2} \right\rceil}$ \\ [1mm]
        $\Pi_{\mathrm{0}}$, $\mathrm{D}_{\mathrm{0}}$ & ${B \times C \times \left\lceil \frac{A_p}{3} \right\rceil \times \left\lceil \frac{L_p}{2} \right\rceil - 1}$ \\ [1mm]
        $\Pi_{\mathrm{1}}$, $\mathrm{D}_{\mathrm{1}}$ & ${B \times 2C \times \left\lceil \frac{A_p}{9} \right\rceil \times \left\lceil \frac{L_p}{4} \right\rceil}$ \\ [1mm]
        $\Pi_{\mathrm{2}}$, $\mathrm{D}_{\mathrm{2}}$ & ${B \times 4C \times \left\lceil \frac{A_p}{9} \right\rceil \times \left\lceil \frac{L_p}{8} \right\rceil}$ \\ [1mm]
        $\mathrm{Y}'_p$ & ${B \times 1 \times \left\lceil \frac{A_p}{3} \right\rceil \times \left\lceil \frac{L_p}{2} \right\rceil - 1}$ \\ [1mm] \hline
    \end{tabular}
\end{table}

% \begin{figure*}[t]
% \centering
% \includegraphics[width=.98\textwidth]{Figures/post_denoiser2.jpg}
% \caption{Pipeline of the dual purpose post-denoiser network with the fusion block.}
% \label{post_denosing_block} 
% \end{figure*}

All the 2D overlapping prediction patches, $\text{Y}'^{(a,l)}_p$, are first obtained within the \(k\)-th region (\(a, l\) are the top-left axial and lateral coordinates of the patches, respectively). Then the reconstruction of that particular region can be produced using 2D weighted windows. The process is described as

\begin{equation}\label{patch_system_0}
\text{Y}'^{a,l}_p= \mathcal{F}_{\text{Recon}}^{\text{patch}}\left(\textbf{I}_{\mathrm{Dp,k}}^{a,l}; \theta_\mathcal{P}^{\text{patch}}\right), \;\; \text{Y}'_p \in \mathbb{R}^{A \times L } 
\end{equation}

\begin{equation} \label{patch_system_1}
\mathbb{P}_k = \begin{bmatrix}
\text{Y}'^{a_1,l_1}_p  & \cdots & \text{Y}'^{a_1,L-1-\frac{L_p}{2}}_p \\
\text{Y}'^{a_2,l_1}_p  & \cdots & \text{Y}'^{a_2,L-1-\frac{L_p}{2}}_p \\
\vdots  & \ddots & \vdots \\
\text{Y}'^{A-1-\frac{A_p}{3},l_1}_p  & \cdots & \text{Y}'^{A-1-\frac{A_p}{3},L-1-\frac{L_p}{2}}_p
\end{bmatrix}_{k} 
\end{equation}

\begin{equation} \label{patch_system_2}
{Y}'_k= \Psi \left(\mathbb{P}_k\right), \;\; k \in \{0,1, ..., N-1\}
\end{equation}
where \(\Psi(\cdot)\) indicates the application of 2D windows (e.g., Tukey windows) to weigh and sum all the patches within the regional ROI. After obtaining the primary reconstruction \(\text{Y}'_k\), equations \eqref{Y_N} and \eqref{YM} can be used to finally obtain the denoised output and the segmentation mask.

% change all the a,l into x,y in contest of the vairables

\begin{figure*}[!t]
\centering
\includegraphics[width=.98\textwidth]{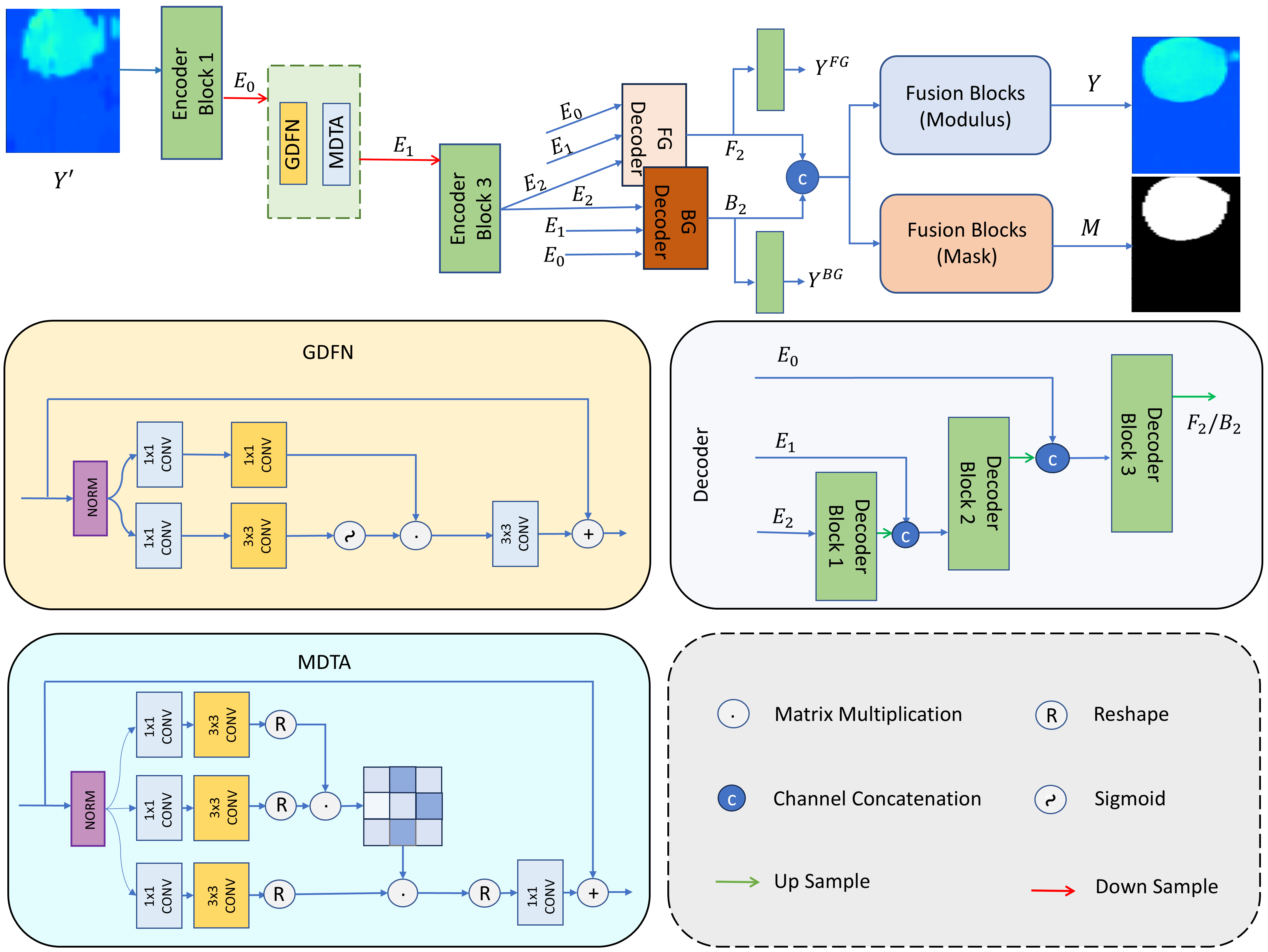}
\caption{Proposed denoiser network architecture. The network encodes input features through an Encoder Block, refines them using GDFN and MDTA modules, and reconstructs foreground and background via separate foreground (FG) and background (BG) Decoders. Final outputs are processed by Modulus and Mask Decoders.}
\label{post_denosing_block} 
\end{figure*}

\subsection{Post Denoiser Network}

\renewcommand{\arraystretch}{1.5}
\begin{table*}[h!]
\centering
\caption{Restormer Architecture Summary}
\label{restomer}

 % Adjust the value to increase or decrease the space between rows

\begin{tabular}{@{}ccccccc@{}}
\toprule
\textbf{Block} & \textbf{Module} & \textbf{Input Resolution}  & \textbf{Modules} & \textbf{Operations} & \textbf{Output} \\ \midrule
\multirow{7}{*}{\rotatebox{90}{Encoder}} & \multirow{3}{*}{$E_0$} & \multirow{3}{*}{$C \times H \times W$}           & Convolution              &  Conv2D ($3 \times 3$)& $H \times W \times C$ \\
                       &                                 &                                                  &   Transformer Block      &   MDTA + GDFN         &    $H \times W \times C$ \\
                       &                                 &                                                  &    Pooling Layer         & $2 \times2$ Maxpool   & $\frac{H}{2} \times \frac{W}{2}$ \\  \cmidrule{2-6}
                       & \multirow{2}{*}{$E_1$}            & \multirow{2}{*}{$ 2C \times\frac{H}{2} \times \frac{W}{2}$}& Transformer Block        & MDTA + GDFN           & $\frac{H}{2} \times \frac{W}{2} \times 2C$ \\
                       &                                 &                                                  &   Pooling Layer           & $2 \times2$ Maxpool  & $\frac{H}{4} \times \frac{W}{4} \times 2C$ \\ \cmidrule{2-6}
                       & \multirow{2}{*}{$E_2$}            & \multirow{2}{*}{$ 2C \times \frac{H}{4} \times \frac{W}{4}$} & Transformer Block        & MDTA + GDFN          & $\frac{H}{4} \times \frac{W}{4} \times 4C$ \\
                       &                                 &                                                   &  Pooling Layer         & $2 \times2$ Maxpool                    & $\frac{H}{8} \times \frac{W}{8} \times 4C$ \\ \midrule
\multirow{8}{*}{\rotatebox{90}{Decoder}} & \multirow{2}{*}{$D_0$}          & \multirow{2}{*}{$ 8C \times \frac{H}{8} \times \frac{W}{8}$}& Transformer Block & MDTA + GDFN & $\frac{H}{8} \times \frac{W}{8} \times 2C$ \\
                         &                               &                                                              & Upsample Layer & $2 \times2$ Upsample   & $\frac{H}{4} \times \frac{W}{4} \times 2C$ \\ \cmidrule{2-6}
                        & \multirow{3}{*}{$D_1$}             & \multirow{3}{*}{$\frac{H}{4} \times \frac{W}{4}$}   & Attention & SE attention & $\frac{H}{4} \times \frac{W}{4} \times 4C$ \\
                         &                                  &                                                     & Transformer Block &  MDTA + GDFN & $\frac{H}{4} \times \frac{W}{4} \times C$ \\ 
                         &                               &                                                              & Upsample Layer & $2 \times2$ Upsample   & $\frac{H}{2} \times \frac{W}{2} \times C$ \\ \cmidrule{2-6}
                         & \multirow{3}{*}{$D_2$}             & \multirow{3}{*}{$\frac{H}{2} \times \frac{W}{2}$}   & Attention & SE attention & $\frac{H}{2} \times \frac{W}{2} \times 2C$ \\
                         &                                  &                                                       & Transformer Block &  MDTA + GDFN & $\frac{H}{2} \times \frac{W}{2} \times C$ \\ 
                         &                                  &                                                         & Upsample Layer & $2 \times2$ Upsample   & $H \times W \times C$ \\ \midrule
          
            \end{tabular}
            \end{table*}

Patch-based training methodologies may introduce spatial inconsistencies, potentially leading to sub-optimal elasticity map  and segmentation mask generations. To address these issues, we propose integrating a post-denoiser block into the model (figure \ref{post_denosing_block}) that both refines the reconstruction output and produces a segmentation mask. The denoising block employs an encoder-decoder architecture (table \ref{restomer}), inspired by the Restormer network \cite{zamir2022restormer}. Our modified decoder block is designed to simultaneously produce a refined reconstruction of SWE image and a segmentation mask.

\subsubsection{Encoder Block} This encoder block aims to compress the primary 2D reconstruction $\text{Y}_{k}'\in \mathbb{R}^{1 \times A \times L}$ in a latent representation to reduce the noise, and preserve the values and structure. Three transformer blocks were used as the feature space encoder. These transformer blocks, $\mathcal{T}(\cdot)$, were composed of a Gated-Dconv feed-forward network (GDFN) block and a multi-Dconv head transposed attention (MDTA) block \cite{zamir2022restormer}. A two-times downsampling was done after each block. The intermediate reconstruction $\text{Y}_k' \in \mathbb{R}^{1 \times A \times L}$ is passed through transformer layers, followed by downsampling operations $\mathrm{P}(\cdot)$, as follows
\begin{equation}
\mathrm{E}_{i} = \mathrm{P}_{2 \times 2}\left(\mathcal{T}_{i}\left(\mathrm{E}_{i-1}; \theta^{t}_{i}\right)\right), \; i=0,1,2
\end{equation} 
Here, $\mathrm{E}_{-1}$ is initialized as $\text{Y}_{k}'$.
% \begin{equation}
%     \mathrm{E}_{0} = \mathrm{P_{2 \times 2}}(\matcal{T_{0}}(\mathrm{Y}_{k}'; \theta^{t}_{0})), \quad \mathrm{E}_0 \in \mathbb{R}^{C \times \frac{L}{2} \times \frac{A}{2}}
% \end{equation}
% This feature space is then passed to the following two stages of the transformer block followed by pooling layers to obtain $\mathrm{E}_{1}$ and $\mathrm{E}_{2}$ as
% \begin{equation}
%     \mathrm{E}_{1} = \mathrm{P_{2 \times 2}}(\matcal{T}_{1}(\mathrm{E}_{0};\theta^{t}_{1})),\quad \mathrm{E}_{1} \in \mathbb{R}^{2C \times \frac{L}{4} \times \frac{A}{4}}
% \end{equation}
% \begin{equation}
%     \mathrm{E}_{2} = \mathrm{P_{2 \times 2}}(\matcal{T}_{2}(\mathrm{E}_{1};\theta^{t}_{2})),\quad \mathrm{E}_{2} \in \mathbb{R}^{4C \times \frac{L}{8} \times \frac{A}{8}}
% \end{equation}

\subsubsection{Lesion Aware Dual Head Decoder Block} 

Our denoiser employs a dual‐decoder design: both decoders share the same encoded feature \(E_{2}\in \mathbb{R}^{8C\times\frac{L}{8}\times\frac{A}{8}}\), but one head refines the Young’s modulus map while the other generates the inclusion segmentation mask. In the first decoding stage, each head upsamples \(E_{2}\) to produce intermediate feature representations \(\mathrm{F}_{0}\) and \(\mathrm{B}_{0}\in \mathbb{R}^{B\times8C\times\frac{L}{4}\times\frac{A}{4}}\), thereby separating lesion and background information into distinct feature spaces. Subsequent decoding stages continue using convolutional padding and fractional interpolation to reconstruct both a denoised elasticity map and a high-confidence segmentation mask, ensuring that residual noise is attenuated in the two outputs and spatial coherence is preserved across patch boundaries. The \(\mathrm{F}_{0}\) and \(\mathrm{B}_{0}\) are then concatenated with $E_1 \in \mathbb{R}^{8C \times \frac{L}{4} \times \frac{A}{4}}$ separately before passing to the second stage of each decoder block where it goes through the similar process. These steps can be described as follows:
\begin{align}
\mathrm{F'}_{i-1} &= \left[\begin{array}{l}\mathrm{F}_{i-1} \\ \mathrm{E}_{2-i}\end{array}\right] \in \mathbb{R}^{4(2-i)C  \times \frac{A}{2(2-i)} \times \frac{L}{2(2-i)}} \\
\mathrm{B'}_{i-1} &= \left[\begin{array}{l}\mathrm{B}_{i-1} \\ \mathrm{E}_{2-i}\end{array}\right] \in \mathbb{R}^{4(2-i)C  \times \frac{A}{2(2-i)} \times \frac{L}{2(2-i)}} \\
\mathrm{F}_{i} &= \operatorname{UP}_{2 \times 2}(\mathcal{F}_{i}(\mathrm{F'}_{i-1}; \theta^{f}_{i})), \nonumber \\
\mathrm{F}_{i} & \in \mathbb{R}^{4(2-i)C  \times \frac{A}{2(2-i)} \times \frac{L}{2(2-i)}} \\
\mathrm{B}_{i} &= \operatorname{UP}_{2 \times 2}(\mathcal{F}_{i}(\mathrm{B'}_{i-1}; \theta^{b}_{i})), \nonumber \\
\mathrm{B}_{i} & \in \mathbb{R}^{(2-i)C  \times \frac{A}{2-i} \times \frac{L}{2-i}}
\end{align}
Two separate transformer blocks are used to estimate foreground (FG) and background (BG) modulus from $ \mathrm{F}_{2} \in \mathbb{R}^{C  \times A \times L}$ and $\mathrm{B}_{2} \in \mathbb{R}^{C  \times A \times L}$ respectively as

% The output from the 2nd stage decoder blocks $\mathcal{F}_{2} \in \mathbb{R}^{C \times A  \times L}$  and $\mathcal{B}_{2}  \in \mathbb{R}^{C  \times A  \times L}$ is extracted following a similar process.

% \begin{align}
%  \mathrm{F'}_{1} &= \left[\begin{array}{l}\mathrm{F}_{1} \\ \mathrm{E}_{2}\end{array}\right], & \mathrm{F'}_{1} & \in \mathbb{R}^{4C  \times \frac{A}{4} \times \frac{L}{4}} \\
%  \mathrm{B'}_{1} &= \left[\begin{array}{l}\mathrm{B}_{1} \\ \mathrm{E}_{2}\end{array}\right], & \mathrm{B'}_{1} & \in \mathbb{R}^{4C \times \frac{A}{4} \times \frac{L}{4}} \\
%  \mathrm{F}_{2} &= \operatorname{UP}_{2 \times 2}( \mathcal{F}_{1}(\mathrm{F'}_{1}; \theta^{f}_{2})), & \mathrm{F}_{2} & \in \mathbb{R}^{C \times \frac{A}{2} \times \frac{L}{2}} \\
%  \mathrm{B}_{2} &= \operatorname{UP}_{2 \times 2}( \mathcal{B}_{2}(\mathrm{B'}_{1}; \theta^{b}_{2})), & \mathrm{B}_{2} & \in \mathbb{R}^{C \times \frac{A}{2} \times \frac{L}{2}}
% \end{align}

\begin{align}
    \mathrm{Y^{FG}} &= \mathrm{ReLU}( {T}_{FG}(\mathrm{F}_{2}, \theta^{t}_{\mathrm{(FG)}})), & \mathrm{Y^{FG}} & \in \mathbb{R}^{A \times L} \\
    \mathrm{Y^{BG}} &= \mathrm{ReLU}( {T}_\mathrm{BG}(\mathrm{B}_{2}, \theta^{t}_{\mathrm{(BG)}})), & \mathrm{Y^{BG}} & \in \mathbb{R}^{A \times L}
\end{align}

 To create the complete denoised reconstruction and a segmentation mask, we merge the feature spaces of the FG and BG, and pass them through two fusion blocks. Each fusion block consists of three sequential transformer blocks that produce the final reconstruction map \({\text{Y} }\) and the segmentation mask \({\text{M}}\) as given by
\begin{align}
{\text{Y} }&= \operatorname{ReLU}\left( \mathcal{F}_{\mathrm{recon}}\left(\begin{bmatrix}\mathrm{B}_2 \\ \mathrm{F}_2\end{bmatrix}; \theta^t_{\mathrm{recon}}\right)\right),  \; {\text{Y}} \in \mathbb{R}^{A \times L} \\
{\text{M}} &= \text{softmax}\left( \mathcal{F}_{\mathrm{seg}}\left(\begin{bmatrix}\mathrm{B}_2 \\ \mathrm{F}_2\end{bmatrix}; \theta^t_{\mathrm{seg}}\right)\right), \;{\text{M}} \in \mathbb{R}^{A \times L}
\end{align}

\subsection{Multi-Task Loss Functions} 
As mentioned in section \ref{introduction}, we designed our network to produce two distinct outputs from the SWE videos: (i) The complete elasticity modulus map of the ROI, (ii) A segmented mask separating the foreground (inclusion) from the background. This multi-task problem requires optimizing two compatible loss functions for each corresponding task. The reconstruction network performs the former tasks. However, the denoising network cleans the modulus mapping, as well as produces a segmentation map (see figure \ref{post_denosing_block}).

\subsubsection{Reconstruction Loss}
To optimize the reconstruction network for producing YM value, we used Mean Absolute Error (MAE) as our first loss function to compute modulus reconstruction loss from the modulus decoder output. The MAE loss $\mathcal{L}_m$ is given by
\begin{equation}
    \label{MAE}
    \mathcal{L}_m=\frac{1}{A L} \sum_{a=0}^{A-1} \sum_{l=0}^{L-1}\left|{\mathrm{y}}_{a, l}^{gt}-\mathrm{y'}_{a, l}\right|
\end{equation}
where ${y}_{a,l}^{gt}$ is the ground truth modulus values and  $y'_{a,l}$ denotes the estimated modulus values at $\left(a, l\right)$ spatial coordinate, $A$ and $L$ represent the number of pixels in the axial and lateral dimensions, respectively.

\subsubsection{Denoising Loss}
In the subsequent stage, we maintain the optimized reconstruction network in a frozen state, and our focus shifts solely to optimizing the weights of the denoiser network. In this phase, our objective is to clean both the FG and BG regions and the fused modulus. To achieve this, we introduce five distinct loss components for the complete denoising task.

(i) \textbf{Foreground (FG) Loss}:  This loss is designed to clean the foreground while ensuring that no modifications are made to the background. It consists of two terms: (a) the foreground loss inside the ground truth inclusion region, denoted as \( \mathcal{L}_{FG1} \),
(b) to further refine the foreground, we also consider regions where the network might overestimate or underestimate the lesion boundary. \( \mathcal{L}_{FG2} \) penalizes false positive predictions by calculating the foreground loss in areas outside the actual lesion, ensuring that the network accurately identifies the relevant structures within the target area. The loss components are as follows:
\begin{equation}\label{FG1_loss}
    \mathcal{L}_{FG1} = \frac{1}{A L} \sum_{a=0}^{A-1} \sum_{l=0}^{L-1} \left| y_{a,l}^{FG} m_{a,l}^{gt} - y_{a,l}^{gt} m_{a,l}^{gt} \right|
\end{equation}
\begin{equation}\label{FG2_loss}
    \mathcal{L}_{FG2} = \frac{1}{A L} \sum_{a=0}^{A-1} \sum_{l=0}^{L-1} \left| y_{a,l}^{FG} (1 - m_{a,l}^{gt}) \right| 
\end{equation}
\begin{equation}\label{FG_loss}
    \mathcal{L}_{FG} = \mathcal{L}_{FG1} + \mathcal{L}_{FG2} 
\end{equation}
(ii) \textbf{Background (BG) Loss }: Similarly, the BG loss aims to clean the background without affecting the foreground. It also includes two terms: (a) the background loss outside the inclusion region, \( \mathcal{L}_{BG1} \), encourages the network to suppress irrelevant details and focus on the region of interest, (b) while \( \mathcal{L}_{BG2} \) addresses false negatives by computing the background loss within the lesion area itself. The loss components are as follows:
\begin{equation}\label{BG1_loss}
    \mathcal{L}_{BG1} = \frac{1}{A L} \sum_{a=0}^{A-1} \sum_{l=0}^{L-1} \left| y_{a,l}^{BG} (1 - m_{a,l}^{gt}) - y_{a,l}^{gt} (1 - m_{a,l}^{gt}) \right| 
\end{equation}
\begin{equation}\label{BG2_loss}
    \mathcal{L}_{BG2} = \frac{1}{A L} \sum_{a=0}^{A-1} \sum_{l=0}^{L-1} \left| y_{a,l}^{BG} m_{a,l}^{gt} \right|
\end{equation}
\begin{equation}\label{BG_loss}
    \mathcal{L}_{BG} = \mathcal{L}_{BG1} + \mathcal{L}_{BG2}
\end{equation}

(iii)  \textbf{Fusion Loss}: To optimize the fusion block, the final output is supervised using the following loss function, similar to the reconstruction loss:
\begin{equation}\label{FUSE_loss}
    \mathcal{L}_{FUSE} = \frac{1}{A L} \sum_{a=0}^{A-1} \sum_{l=0}^{L-1} \left| y_{a,l}^{gt} - y_{a,l} \right|
\end{equation}
The three losses are described in equations (\ref{FG_loss}), (\ref{BG_loss}), and (\ref{FUSE_loss}) clean the regions and merge them optimally on average. 

(iv) \textbf{NCC-loss:} To supervise structural and shape-oriented similarity between the ground truth and predictions, an NCC-loss is used:
\begin{equation}
    \mathcal{L}_{NCC}= 1.0-M_{NCC}
\end{equation}
where,
\begin{equation}
    M_{NCC}= \frac{\sum_{n=0}^{A-1}\sum_{m=0}^{L-1} y^{gt}_{a,l}\cdot y_{a,l}}{\sqrt{\sum_{n=0}^{A-1}\sum_{m=0}^{L-1} (y^{gt}_{a,l})^2 \cdot \sum_{n=0}^{A-1}\sum_{m=0}^{L-1} (y_{a,l})^2} + \varepsilon}
\end{equation}
A small positive number $\varepsilon$ is used to avoid division by zero.

(v) \textbf{TV-loss:} If some impulsive or outlier noise exists in $Y'$ prevailing within 1-2 pixels, then that will not be easily reduced in the MAE sense. As such, an additional loss is considered to smoothen the reconstruction and prevent any sudden noise jumps in both the axial ($a$) and lateral ($l$) directions, namely Total-Variation (TV) loss:
\begin{equation}\label{tv}
    \mathcal{L}_{TV}= 2\;(\mathcal{L}_{TV}^a+\mathcal{L}_{TV}^l)
\end{equation}
where the axial and lateral components are defined as
\begin{equation}\label{tva}
    \mathcal{L}_{TV}^a = \frac{1}{A- 1} \sum_{a=0}^{A - 1} \sum_{l=0}^{L} \left( y_{a,l} - y_{a+1,l} \right)^2
\end{equation}
\begin{equation}\label{tvl}
    \mathcal{L}_{TV}^l = \frac{1}{L - 1} \sum_{a=0}^{A} \sum_{l=0}^{L - 1} \left( y_{a,l} - y_{a,l+1} \right)^2
\end{equation}

We combine the losses in equations (\ref{FG_loss}), (\ref{BG_loss}), (\ref{FUSE_loss}), and (\ref{tv}) with appropriate coupling coefficients to form the total denoising loss, $\mathcal{J}_{denoise}$ :

\begin{equation}\label{denoise}
    \mathcal{J}_{Denoise} = \alpha \mathcal{L}_{FG} +\beta \mathcal{L}_{BG}+ \lambda \mathcal{L}_{FUSE} + \eta \mathcal{L}_{NCC}+ \gamma \mathcal{L}_{TV}
    \end{equation}
Here, $\alpha$ and $\beta $ are dynamically selected by the mean ratio between BG and FG pixels for each data, $\delta$ and $\gamma$ are selected iteratively. The $L_{FUSE}$ coefficient contains both the FG and BG-loss coefficients as it deals with the entire region.

(vi) \textbf{Segmentation loss}: To optimize our denoiser network to produce a segmentation mask from the input data, we used the \textbf{IoU} Loss function to optimize the output of the mask decoder. This helps the network output to focus on isolating the predicted inclusion position with the ground truth inclusion position. If $M^{gt}$ is defined as a ground truth binary mask and $M$ is defined as a predicted binary mask, the \textbf{IoU} function is defined as 

\begin{equation}
    \label{jaccard}
    J(M^{gt}, M)
    =\frac{|M^{gt} \cap M|}{|M^{gt}-M|+|M^{gt} \cap M|+|M-M^{gt}|+\varepsilon}
\end{equation}
Then the \textbf{IoU} Loss can be defined as
\begin{equation}
    \label{IoULoss}
    \mathcal{L}_{IoU}=1.0-J(M^{gt}, M) %=1-\frac{|M \cap \hat{M}|}{|M \cup \hat{M}|+ c}
\end{equation}
We combine the losses in equations (\ref{denoise}) and (\ref{IoULoss}) using a coupling factor $\mu$ to obtain a comprehensive total loss as

\begin{equation}\label{joint_loss}
    \mathcal{J}_{total} =\mathcal{J}_{denoise} +\mu \mathcal{L}_{\mathrm{IoU}}(M^{gt},M)
\end{equation}
The total loss in equation (\ref{joint_loss}) is used to optimize the proposed denoiser network for both cleaning and mask generation simultaneously. 
% In the initial stage, we disconnect the denoising network and focus on optimizing the output of the reconstruction network, which includes variables such as $Y^'$ and $M$, alongside the ground truth data $Y^{gt}$ and $M^{gt}$. The reconstruction loss, denoted as $\mathcal{J}_{rec}$, drives this optimization process.

\section{Experimental Setup}

\subsection{Simulation of Shear Wave Propagation}
\label{dataset_sim}
To generate tissue displacement data in a simulation environment, we employed COMSOL Multiphysics’s Structural Mechanics module, as recommended in prior studies \cite{6264136,ahmed2021dswe}. The acoustic radiation force (ARF) was modeled using a Gaussian axial force distribution (see Eq.~\ref{ARF}) to induce a propagating shear wave within a 3D finite‐element model (FEM).

\begin{equation}\label{ARF}
{ARF}= A \exp \left\{-\left(\frac{\left(x-x_0\right)^2}{2 \sigma_x^2}+\frac{\left(z-z_0\right)^2}{2 \sigma_z^2}\right)\right\}
\end{equation}
Here, \((x_{0}, z_{0})\) denotes the ARF focus point in the lateral and axial directions, and \(\sigma_{x}\) and \(\sigma_{y}\) are the beam widths in those respective axes. The amplitude \(A\) is set to \(1000\ \mathrm{Nm^{-3}}\), and the push duration is \(400\,\mu\text{s}\), ensuring that the maximum tissue displacement does not exceed \(20\,\mu\text{m}\), in accordance with safety limits on mechanical and thermal indices.
 
In contrast to methods that generate elasticity maps from a single push outside the ROI \cite{sarvazyan1998shear} or that require multiple pushes (e.g., CUSE \cite{song2012comb,song2014fast} and supersonic shear imaging (SSI) \cite{bercoff2004supersonic}), our approach divides the ROI into \(N\) overlapping subregions. Displacement data for each subregion are acquired via a separate ARF push, positioned 4 $mm$ lateral to the target zone—after which individual elasticity estimates are computed and fused using tapered‐window weighting. Thus, estimating the elasticity of a single phantom entails \(N\) sequential pushes.

\begin{table}[h]
\centering
\caption{Simulation parameters for shear wave generation}
\begin{tabular}{cc}
\hline
\textbf{Parameters} & \textbf{Value} \\
\hline
ARF intensity, A & $2 \times 10^5 \ \mathrm{N} / \mathrm{m}^3$ \\
\hline
$\sigma_x, \sigma_y$ & $0.44 \ \mathrm{mm}, 8.00 \ \mathrm{mm}$ \\
\hline
$x_0, z_0$ & $variables$ \\
\hline
\multirow{2}{*}{Medium} & Nearly incompressible \\
 & linear, isotropic, elastic solid \\
\hline
Poisson's ratio, $v$ & 0.499 \\
\hline
Density, $\rho$ & $1000 \ \mathrm{kg} / \mathrm{m}^3$ \\
\hline
ARF excitation time & $400 \ \mu \mathrm{s}$ \\
\hline
Wave propagation time & $8 \ \mathrm{ms}$ \\
\hline
FEM size & $38\ \mathrm{mm} \times 40 \ \mathrm{mm}$ \\
\hline
Mesh element & Triangular \\
\hline
\end{tabular}

\end{table}
A low‐reflecting boundary condition was applied at the edges of the field of view in our simulated phantoms to suppress residual reflections and minimize artifacts. Such conditions were not imposed at inclusion boundaries, since eliminating reflections in vivo is not feasible. We generated 1,380 bi‐level phantom datasets using the COMSOL–MATLAB interface, randomly varying inclusion diameter, inclusion position, and both inclusion and background stiffness. These datasets were partitioned into training, validation, and test sets. As described in Section \ref{methods}, an \(N\)-push imaging sequence was employed to cover the full ROI: each push targeted one of \(N\) lateral positions, producing \(N\) overlapping subregions whose elasticity estimates were later fused.

In the simulation environment, the following procedure was employed:

\begin{itemize}
    \item \textbf{ROI Selection}: A region of interest (ROI) measuring $17.5\ \text{mm} \times 25.7\ \text{mm}$ was fixed within a field of view (FOV) of $38\ \text{mm} \times 40\ \text{mm}$.
    
    \item \textbf{Inclusion Generation}: Circular inclusions with diameters uniformly sampled between $3\ \text{mm}$ and $12\ \text{mm}$ were placed at random positions within the ROI.
    
    \item \textbf{Stiffness Variation}: Inclusion stiffness values were drawn randomly from $8\ \text{kPa}$ to $100\ \text{kPa}$, and background stiffness values from $10\ \text{kPa}$ to $35\ \text{kPa}$.
    
    \item \textbf{Imaging Sequences}: For each ROI, $N$ ARF pushes were applied, dividing the ROI into $N$ equal subregions. Each push was offset by $4\ \text{mm}$ laterally.
    
    \item \textbf{Resolution Settings}: The axial sampling density was set to $8\ \text{pixels/mm}$ and the lateral density to $0.7\ \text{pixels/mm}$, yielding subregion dimensions of $20.5\ \text{mm} \times 7\ \text{mm}$.
    
    \item \textbf{Frame Rate}: Shear wave tracking was performed at a pulse repetition frequency of $8\ \text{kHz}$ (i.e., $125\ \mu\text{s}$ between frames).
\end{itemize}

For all simulations, we selected $N=4$ imaging sequences.

\subsection{CIRS049 Phantom Dataset}
\label{dataset_cirs}

A total of 72 SWE imaging cases were acquired from Fuji Healthcare (USA) using CIRS049 phantoms labeled A, B, C, and D, each containing inclusions of stiffness Types 1–4. From each phantom, 18 ROIs were selected for SWE data collection. A summary of the collected CIRS049 phantom dataset is presented in Table~\ref{CIRS049}.

\begin{table}[!h]
    \centering
    \caption{Description of CIRS phantom data type. The types 1-4 indicate the foreground (FG).}
    \begin{tabular}{c|ccccc}
    \hline
        CIRS049 & \multicolumn{5}{c}{Type (kPa)} \\ \cline{2-6}
         & BG & 1 & 2 & 3 & 4 \\ \hline
        A & 24 & 7 & 12 & 39 & 66 \\ 
        B & 21 & 6 & 9 & 36 & 76 \\ 
        C & 18 & 6 & 9 & 36 & 72 \\ 
        D & 20 & 6 & 9 & 36 & 72 \\ \hline
    \end{tabular}
    \label{CIRS049}
\end{table}

Four imaging sequences were performed per ROI, each covering an overlapping subregion. The lateral push location remained fixed relative to the ROI, while the axial push position varied to target inclusions at different depths, yielding $N=4$ subregions per dataset.

The acquired raw in‐phase and quadrature (IQ) data were preprocessed to enhance signal quality. The resulting displacement fields, indicative of tissue elasticity, were then extracted for subsequent analysis.

\subsection{Training Procedure}
\subsubsection{Implementation Details: Simulation Data }
\label{train_sim}

To ensure robustness over a range of inclusion stiffnesses, we partitioned the simulated COMSOL dataset into nonoverlapping training, validation, and test sets by randomly assigning stiffness values. This resulted in 1,010 ROIs (each split into four subregions) for training, 111 for validation, and 259 for testing. Each subregion measured $168\times10$ pixels ($25.7\ \mathrm{mm}\times7\ \mathrm{mm}$) with 72 time frames stacked to form a three‐dimensional displacement volume. We interpolated the lateral dimension to 16 pixels, yielding input volumes of size $72\times168\times16$ (time $\times$ axial $\times$ lateral). For the CIRS phantom data, 70\% of samples were allocated to training. Ground‐truth stiffness maps were exported from the COMSOL–MATLAB interface, and binary inclusion masks were generated by thresholding.

All modulus maps were normalized by mapping the maximum stiffness to 100~kPa, and displacement volumes were scaled using min–max normalization. The network was trained on an NVIDIA RTX 4090 GPU with a batch size of 16, using the Adam optimizer with an initial learning rate of $10^{-4}$. A learning‐rate scheduler reduced the rate by a factor of 0.2 after two epochs without validation‐loss improvement. Training continued for 150 epochs to ensure convergence on both training and validation sets.

\subsubsection{Implementation Details: CIRS049 Data}
\label{Implementation_CIRS}

For the CIRS049 phantom (Section \ref{dataset_cirs}), we split data into 70\% training, 15\% validation, and 15\% test sets. To augment the limited phantom samples, we combined its training portion with the simulated COMSOL training data. Ground-truth elasticity maps for CIRS049 were generated by manually drawing a binary mask on the B-mode phantom image and assigning modulus values as specified in table \ref{CIRS049}. All other training parameters, including normalization, optimizer settings, learning-rate schedule, and epoch count, remained identical to those used for the simulation data (Section \ref{train_sim}).

% \begin{figure*}[!t]
% \centering
% \includegraphics[width=.8\textwidth]{Figures/Patch 2.png}
% \caption{Scemetic diagram for Patch-Based Modulus Creation}
% \label{post_denosing_block} 
% \end{figure*}
% \subsubsection{Patch Implementation on CIRS Phantom Data}
% The total training sample for CIRS phantom data is 181 which is insufficient for training the neural network with complex and diverse displacement data. To better generalize the neural network we implemented an overlapping patch-based training approach to multiply the limited data to a large number. For this, we divided the SWEI into 154 overlapping patches. The axial and lateral dimensions of this patch are $21 \times 4$. To estimate this patched from the shear wave displacement data, A window shape of $63 \times 10$ was taken for input of the reconstruction network. The overlapped region in lateral direction was 2 pixels and the overlap region for axial dimensions was 14 pixels concerning the ground truth.

\subsection{Evaluation Metrics}
We performed a quantitative evaluation of our proposed network using four quality metrics for the reconstruction task: (i) Peak Signal-to-Noise Ratio (PSNR), (ii) Contrast-to-Noise Ratio (CNR), (iii) Structural Similarity Index (SSIM), and (iv) Mean Absolute Error (MAE). For the segmentation task, we used four additional quality metrics: (i) Intersection over Union (IoU), (ii) F1-Score, (iii) Hausdorff Distance (HD), and (iv) Average Symmetric Surface Distance (ASSD).
\vspace{3mm}

\textbf{PSNR}: Peak Signal-to-Noise Ratio is a commonly used index to measure the reconstruction quality of an image. It is defined as 
\begin{eqnarray}
    \label{psnr}
    \mathrm{PSNR}  &=&  10 \cdot \log _{10}\left(\frac{1}{\mathrm{MSE}}\right)
\end{eqnarray}
where $\mathrm{MSE}$ is calculated as
\begin{eqnarray}
\text{MSE} &=& \frac{1}{mn} \sum_{i=0}^{m-1} \sum_{j=0}^{n-1} \left[I_{\text{norm}}^{(i, j)} - \hat{I}_{\text{norm}}^{(i, j)}\right]^2 \\
I_{\text{norm}} &=& \frac{I}{\text{max}(I)}, \quad \hat{I}_{\text{norm}} = \frac{\hat{I}}{\text{max}(\hat{I})}
\end{eqnarray}
Here, $\mathcal{I}$ and $\hat{\mathcal{I}}$ are the ground-truth 2D modulus mapping and the estimated 2D modulus mapping, respectively.
\vspace{3mm}

\textbf{CNR}: Contrast to Noise Ratio is an image quality metric calculated as in the following equation.
\begin{equation}
    \label{CNR}
    \mathrm{CNR}=10 \log _{10}\left(\frac{|\mu_I-\mu_B|}{\sigma_B^2}\right)
\end{equation}

where $\mu_I$, $\mu_B$ is the mean elasticity value of the inclusion and background, respectively and $\sigma_B^2$ is the variance of the elasticity of the background
\vspace{3mm}

\textbf{SSIM}:  Structural Similarity Index is a quantitative analysis of the perceived quality between two images. It is denoted as     
\begin{equation}
    \label{SSIM}
  \mathrm{SSIM}(I, \hat{I}) = \frac{{(2\mu_I\mu_{\hat{I}} + C_1)(2\sigma_{\mathrm{cov}} + C_2)}}{{(\mu_I^2 + \mu_{\hat{I}}^2 + C_1)(\sigma_{I}^2 + \sigma_{\hat{I}}^2 + C_2)}} 
\end{equation}
where $\mu_I$ and  $\mu_{\hat{I}}$ represent the mean of the original and reconstructed image, respectively, $\sigma_{I}$ and  $\sigma_{\hat{I}}^2$ represent the standard deviation of the original and reconstructed image, respectively, $\sigma_{\mathrm{cov}}$ represents the covariance between them. The constants $C_1$ and $C_2$ are used to avoid division by zero.

\textbf{MAE}: The Mean Absolute Error (MAE), a metric used to assess the quality of elasticity reconstruction, is calculated separately for the FG and BG regions to provide a more comprehensive evaluation of the reconstruction's accuracy. It was defined in equation (\ref{MAE}).

\textbf{IoU Score}: It is used to quantify the network performance in the segmentation task. The IoU or Jaccard metric was mentioned in equation (\ref{jaccard})

% . By taking (1-$L_{IoU}$), the performance metric is obtained: 
% \begin{equation}
%     \label{IoU_metric}
%   \mathrm{IoU}(\textbf{M}^{gt}, \textbf{M}) = \frac{|\textbf{M}^{gt} \cap \textbf{M}|}{|\textbf{M}^{gt} \cup \textbf{M}|+\varepsilon}
% \end{equation}

\textbf{F1-Score:} The F1-score is calculated from the harmonic mean between the precision and recall metrics from any prediction. We utilize this metric for observing our segmentation performance. It can also be calculated directly using true-positive (TP), false-positive (FP) and false-negative (FN) values as
\begin{equation}
    \label{F1-score}
     \mathrm{F1} = \frac{2TP}{2TP+FP+FN}
\end{equation}

\begin{table*}[t]
\centering
\renewcommand{\arraystretch}{1}
\caption{Quantitative Comparison among the test cases from different Datasets [$\uparrow$: higher is better, $\downarrow$: lower is better]}
\label{Test cases result comparison}
\begin{tabular}{ccccccccc}
    \hline
    \textbf{Data} & \textbf{Method} & \textbf{MAE}$\downarrow$ & \textbf{MAE}$\downarrow$ & \textbf{CNR}$\uparrow$ & \textbf{PSNR}$\uparrow$ & \textbf{PSNR}$\uparrow$ & \textbf{PSNR}$\uparrow$ & \textbf{SSIM}$\uparrow$ \\
     &  & \textbf{(FG)} & \textbf{(BG)} &  &  & \textbf{(FG)} & \textbf{(BG)} &  \\
     &  & \textbf{[kPa]} & \textbf{[kPa]} & \textbf{[dB]} & \textbf{[dB]} & \textbf{[dB]} & \textbf{[dB]} &  \\
    \hline
    \multirow{4}{*}{} & DSWENet \cite{ahmed2021dswe} & 3.87 & 2.32 & 29.30 & 21.28 & 18.04 & 22.49 & 0.932\\
     \textbf{Simulation} & Neidhardt et al. \cite{neidhardt2022ultrasound} & 0.76 & 0.15 & 41.04 & 29.09 & 27.78 & 29.48 & 0.996 \\
    (SNR: $\infty$ dB)& \textbf{Ours} ($\mathrm{Y}'$) & 1.37 & 0.27 & 40.53 & 29.14 & 25.02 & 32.34 & 0.989 \\
    & \textbf{Ours} ($\mathrm{Y}$) & 0.92 & 0.11 & 46.93 & 33.96 & 28.09 & 32.46 & 0.997 \\
    \hline
    \multirow{3}{*}{} & DSWENet \cite{ahmed2021dswe} & 4.67 & 2.45 & 25.84 & 19.17 & 16.08 & 21.33 & 0.911 \\
     {\textbf{Simulation}}& Neidhardt et al. \cite{neidhardt2022ultrasound} & 3.22 & 1.03 & 28.97 & 19.96 & 20.02 & 19.32 & 0.932 \\
    % & \textbf{Ours} ($\mathrm{Y}'$) & 1.84 & 0.24 & 39.83 & 28.89 & 24.78 & 28.47 & 0.991 \\
    (SNR: 11 dB)& \textbf{Ours} ($\mathrm{Y}$) & 1.42 & 0.16 & 46.78 & 32.68 & 26.09 & 31.21 & 0.995 \\
    \hline
    \multirow{3}{*}{CIRS049} & DSWENet \cite{ahmed2021dswe} & 8.51 & 2.40 & 26.06 & 15.96 & 12.92& 15.91 & 0.899 \\
     & Neidhardt et al. \cite{neidhardt2022ultrasound} & 8.00 & 1.31 & 25.29 & 16.30 & 12.08 & 16.09 & 0.918 \\
    % & \textbf{Ours} ($\mathrm{Y}'$) & 8.30 & 1.55 & 28.33 & 15.73 & 19.45 & 19.71 & 0.910 \\
    % & \textbf{Ours (Patched)} & 6.09 & 1.31 & 28.91 & 19.75 & 14.61 & 16.07 & 0.901 \\
    % & \textbf{Ours ($\mathrm{Y}$)} & 5.34 & 1.34 & 38.21 & 20.83 & 14.90 & 17.30 & 0.933 \\
    & \textbf{Ours ($\mathrm{Y}$)} & 5.68 & 0.99 & 42.14 & 21.11 & 15.58 & 17.84 & 0.936 \\
    \hline
\end{tabular}
\end{table*}

\textbf{Hausdorff Distance (HD) \cite{huttenlocher1993comparing}:} The HD index determines the maximum distance between two sets of surfaces. The differences between two sets of surface pixels are calculated using Euclidean distance and the maximum value results in the HD index. If we assume $\Omega(M^{gt})$ and $\Omega(M)$ to be the surface pixels of the ground and prediction mask, respectively, the metric can be defined as
\begin{equation}
    \begin{aligned}
        \mathrm{HD}\;(\textbf{M}^{gt}, \textbf{M}) =  max & \bigg \{   \max_{s_1 \in \Omega(\textbf{M}^{gt})} d(s_1,\Omega(\textbf{M})), \\
        & \max_{s_2 \in \Omega(\textbf{M})} d(s_2,\Omega(\textbf{M}^{gt})) \bigg \}        
    \end{aligned}     
\end{equation}
where,
\begin{equation}
    d(v, \Omega(W)) = \min_{s_w \in \Omega(W)} \sqrt{\sum_{i=1}^n (v_i - s_{w,i})^2}
\end{equation}
Here, \( v = (v_1, v_2, \ldots, v_n) \) and \( s_w = (s_{w,1}, s_{w,2}, \ldots, s_{w,n}) \) are vectors in \( \mathbb{R}^n \).

\textbf{Average Symmetric Surface Distance \cite{heimann2009comparison}:} HD finds the worst-case scenario of distances between the two sets of mask surfaces. We can also use the average symmetric surface distance (ASSD) to determine the overall mask surface gaps in the test cases. Being a minimizing metric, it uses the number of pixels, $n(M^{gt})$ and $n(M)$, of the masks to normalize cross-surface distances. ASSD is calculated as
\begin{equation}
    \begin{aligned}
        \mathrm{ASSD}\;(\textbf{M}^{gt}, \textbf{M}) = \frac{1}{\mathcal{N}}& \bigg [   \sum_{s_1 \in \Omega(\textbf{M}^{gt})} d(s_1,\Omega(\textbf{M})) \\
        & + \sum_{s_2 \in \Omega(\textbf{M})} d(s_2,\Omega(\textbf{M}^{gt})) \bigg ]        
    \end{aligned}     
\end{equation}
where,
\begin{equation}
    \mathcal{N} = n(\textbf{M}^{gt}) + n(\textbf{M})
\end{equation}
Both HD and ASSD are minimizing metrics.

\section{Results}
In this section, we present a thorough evaluation of our proposed method against previously reported deep learning techniques (Ahmed et al. \cite{ahmed2021dswe} and Neidhardt et al. \cite{neidhardt2022ultrasound}) using both finite element phantom data and experimental CIRS049 phantom data. This assessment encompasses both qualitative visual comparisons and quantitative analysis of performance metrics.

\subsection{Performance Evaluation: Simulation Data}\label{Simu_performance}

To rigorously assess the robustness and accuracy of our deep learning pipeline, we conducted a series of evaluations on simulated stiffness maps.
 We generated these maps under two conditions: using noise-free simulated displacement data and introducing additive Gaussian noise to the motion data, achieving 11 dB SNR.  Our pipeline was trained using the methodology outlined in equations (\ref{Y}) and (\ref{YM}).

Quantitative results, presented in table \ref{Test cases result comparison}, reveal notable performance differences between the evaluated methods. DSWENet \cite{ahmed2021dswe} demonstrates substantially higher FG and BG MAE (FG $3.87$~kPa; BG $2.32$~kPa) in the noiseless scenario compared to the method by Neidhardt et al. \cite{neidhardt2022ultrasound}. Neidhardt et al.'s method achieves low FG MAE ($0.76$~kPa) and BG MAE ($0.15$~kPa) on clean simulated data outperforming our primary reconstruction (FG $1.37$~kPa, BG $0.27$ kPa). Our denoised outputs exhibit a notable decrease in both FG and BG MAEs (FG $0.92$~kPa, BG $0.11$~kPa), achieving almost equivalent performance to that of Neidhardt et al. \cite{neidhardt2022ultrasound}.

Our primary reconstruction model performs well on clean data, achieving a CNR of $40.53$~dB and a PSNR of $29.14$~dB. While Neidhardt et al.~\cite{neidhardt2022ultrasound} results in slightly higher CNR ($41.04$~dB) while PSNR is lower ($29.09$~dB) which indicates similar smoothness in both methods. Our denoised model improves PSNR in FG to $28.09$~dB, compared to $27.78$~dB for Neidhardt et al., though with a slightly higher FG MAE ($0.92$ vs. $0.76$). A similar trend is observed in the background (BG): Neidhardt et al. achieve a lower BG MAE ($0.15$~kPa) than our main model ($0.27$~kPa), but our denoised version performs best with $0.11$~kPa. For BG PSNR, both our models outperform Neidhardt et al., with values of $32.34$~dB (primary) and $32.46$~dB (denoised). After denoising, our overall CNR and PSNR improve to $46.93$~dB and $33.96$~dB, respectively, surpassing Neidhardt et al.'s results. SSIM is comparable between our denoised model ($0.996$) and Neidhardt et al. ($0.997$), while DSWE-Net performs noticeably worse with an SSIM of $0.932$.

Quantitative analysis of the $11$~dB SNR simulation, summarized in Table~\ref{Test cases result comparison}, highlights the effects of noise on reconstruction performance. Neidhardt et al.~\cite{neidhardt2022ultrasound} shows increased error under noise, with FG and BG MAE rising to $3.22$~kPa and $1.03$~kPa, respectively. DSWENet~\cite{ahmed2021dswe} is slightly more robust but still performs poorly, yielding FG MAE of $4.67$~kPa and BG MAE of $2.45$~kPa. In contrast, our method maintains strong performance with FG MAE of $1.42$~kPa and BG MAE of $0.16$~kPa, approximately $3.25\times$ and $6.44\times$ lower than DSWENet, respectively. Moreover, our method reaches $46.78$~dB CNR, outperforming DSWENet ($25.84$~dB) and Neidhardt et al. ($28.97$~dB), showing a $17.81$~dB gain over the latter. PSNR results are similarly strong, with our model achieving $32.68$~dB overall, compared to $19.17$~dB (DSWENet) and $19.96$~dB (Neidhardt et al.). Our BG PSNR of $31.21$~dB also exceeds that of DSWENet ($21.33$~dB) and Neidhardt et al. ($19.32$~dB), indicating effective noise suppression in homogeneous regions. The FG PSNR of $26.09$~dB is higher than both DSWENet ($16.08$~dB) and Neidhardt et al. ($20.02$~dB). Finally, our method achieves near-perfect (SSIM $= 0.995$), surpassing DSWENet ($0.911$) and Neidhardt et al. ($0.932$). The $0.063$ SSIM improvement over Neidhardt et al. reflects better preservation of anatomical detail under noise.

 \begin{figure*}[t]
  \centering
  \includegraphics[width=0.8\textwidth]{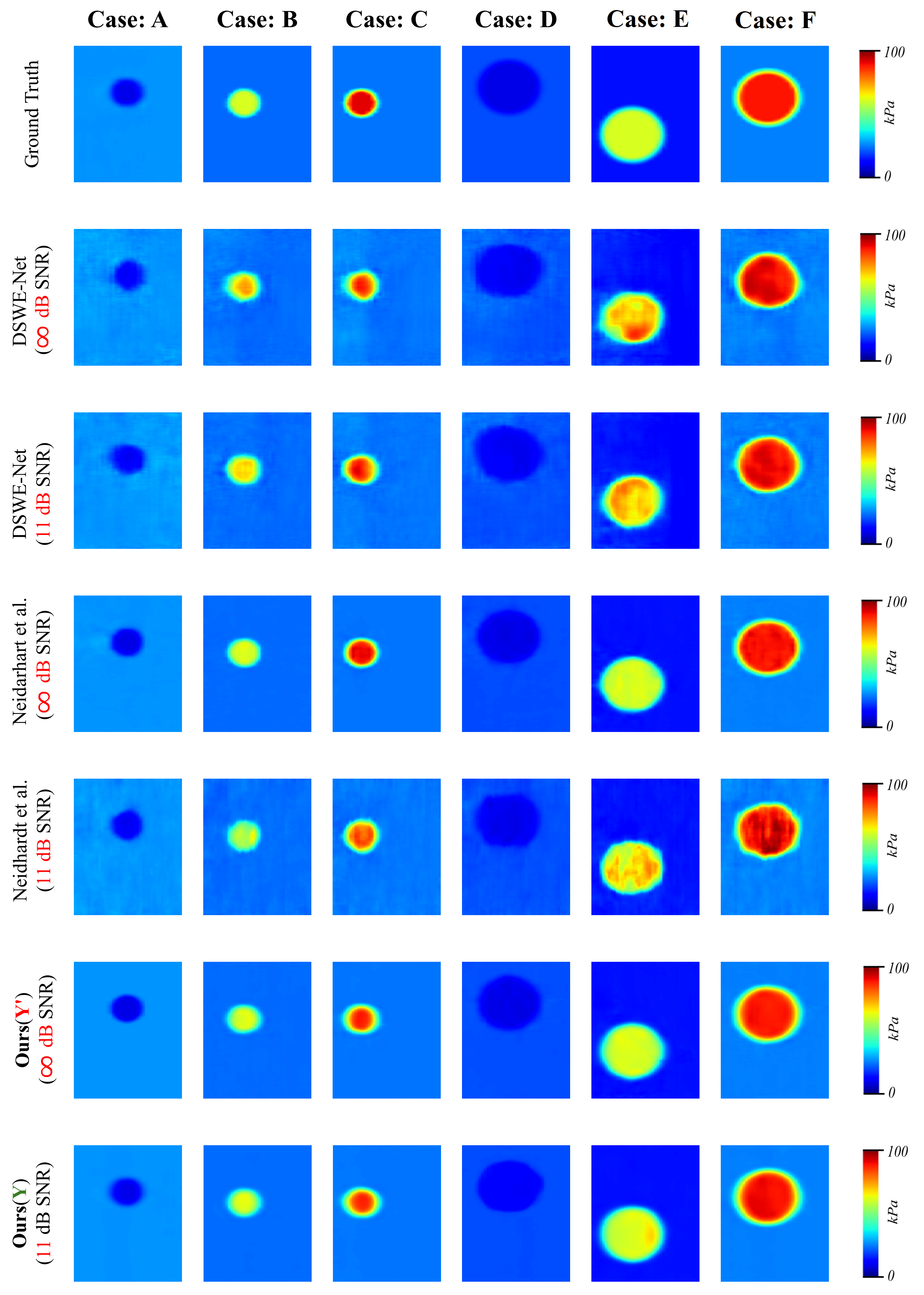}
  \caption{Comparison of ground truth and 2D YM map of various image reconstruction methods across different cases (A to F). The top row shows the ground truth images for each case. The subsequent rows display the reconstructed YM map using DSWE-Net \cite{ahmed2021dswe}, Neidhardt et al.'s \cite{neidhardt2022ultrasound} method, and our method at different SNR levels ($\infty$ dB and 11 dB). The color scale on the right indicates the kPa range.}
  \label{result_comp_simu} 
\end{figure*}

\begin{figure*}[t]
  \centering
  \includegraphics[width=0.8\textwidth]{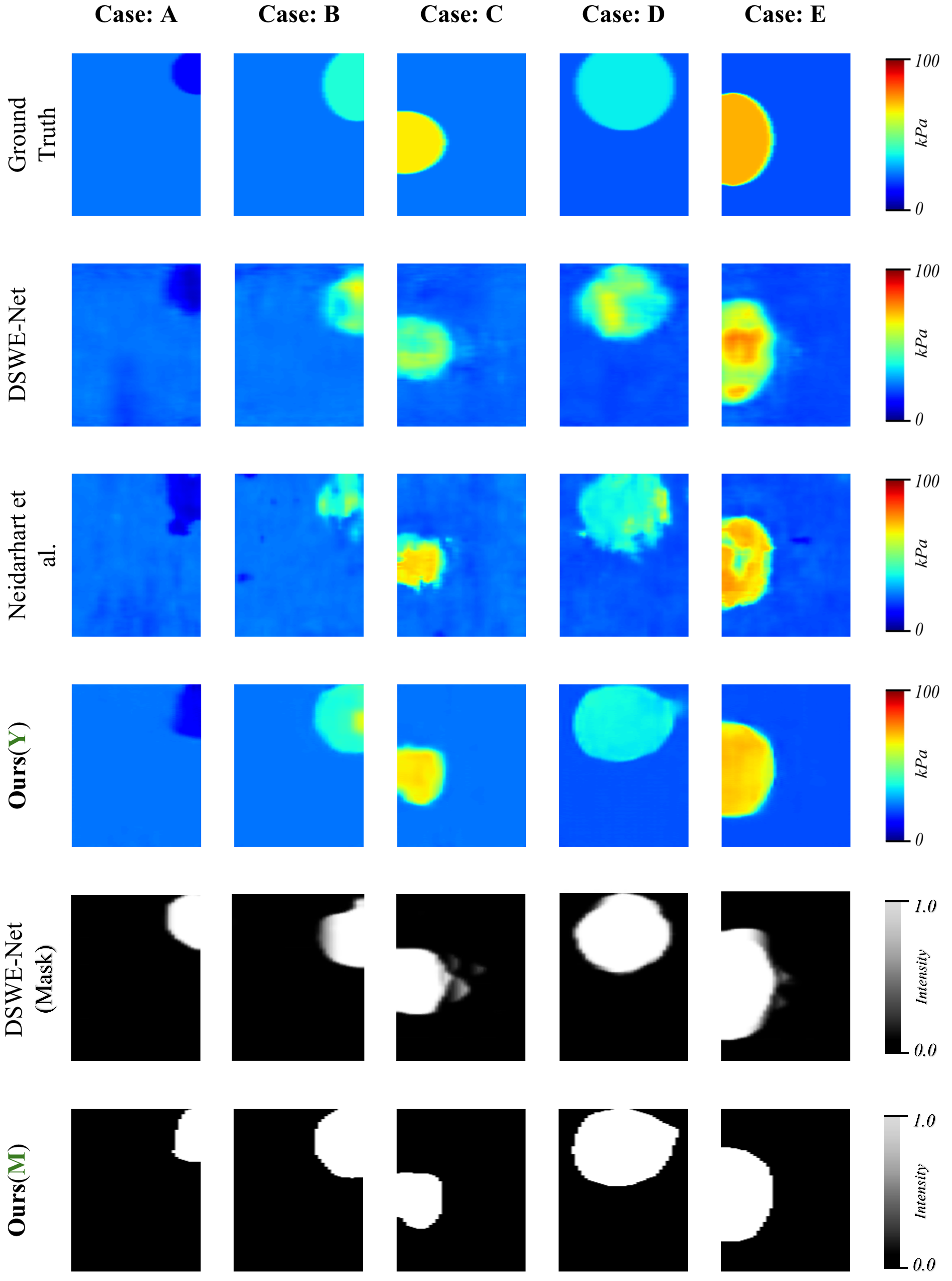}
  \caption{2-D YM image reconstruction for the CIRS049 test phantoms. The top row shows the ground truths of the five separate phantoms (Case: A-E). And, the next rows show the results from DSWE-Net \cite{ahmed2021dswe}, Neidahart et al. \cite{neidhardt2022ultrasound}, and denoised output from our method, respectively. The segmentation results are from DSWE-Net and our pipeline.}
  \label{result_comp_cirs} 
\end{figure*}

\subsection{Performance Evaluation: CIRS049 Data}\label{CIRS_performance}

Evaluation on the CIRS049 phantom (Table~\ref{Test cases result comparison}) confirms our method’s clinical relevance. DSWENet~\cite{ahmed2021dswe} shows high FG MAE ($8.51$~kPa) and moderate BG MAE of ($2.40$~kPa), while Neidhardt et al.~\cite{neidhardt2022ultrasound} reports FG/BG MAE to be $8.00$/$1.31$~kPa. Our approach reduces FG MAE to $5.68$~kPa (a $33.3\%$ decrease) and BG MAE to $0.99$~kPa (a $58.8\%$ decrease) relative to Neidhardt et al.

For CNR, we achieve $42.14$~dB, which exceeds that of DSWENet by $16.08$~dB and Neidhardt’s $25.29$~dB by $16.85$~dB, indicating substantially better lesion visibility. Our PSNR of $21.11$~dB also surpasses DSWENet ($15.96$~dB) and Neidhardt ($16.30$~dB), with BG PSNR of $17.84$~dB versus $15.91$/$16.09$~dB and FG PSNR of $15.58$~dB versus $12.92$/$12.08$~dB. Although Neidhardt’s point-wise estimation uses all data, its noise sensitivity yields the lowest CNR and PSNR. The SSIM of $0.936$ outperforms DSWENet ($0.899$) and Neidhardt ($0.918$), reflecting superior structural fidelity via our patch-based training (Section~\ref{Patch}).

% In contrast, our localized, denoising-guided paradigm (Eqs.~\ref{Y_N}–\ref{YM}) effectively learns from limited phantom data while maintaining robustness to measurement noise.

% SSIM of $0.936$ of our method exceeds both DSWENet ($0.899$) and Neidhardt et al. ($0.918$). This $0.018$ SSIM improvement over Neidhardt et al., achieved through our patch-based training strategy (Section~\ref{Patch}), reflects better preservation of anatomical structures despite limited training data and inherent measurement noise.
% Notably, while Neidhardt et al. benefits from exhaustive data utilization through point-wise estimation, its poor noise handling results in the lowest CNR ($25.29$~dB) and PSNR ($16.30$~dB) values. Our patch-based paradigm, guided by Equations~(\ref{Y_N}), (\ref{patch_system_1}), (\ref{patch_system_2}), and (\ref{YM}), demonstrates that localized processing combined with denoising enables effective learning from scarce phantom data while maintaining noise robustness.

\subsection{Performance Evaluation: Segmentation Mask } \label{mask_performance}

\begin{table}[t]
    \small
    \centering
    \caption{Segmentation performance on test cases of different datasets [$\uparrow$: higher is better, $\downarrow$: lower is better]}
    \label{table:Masks}
    \resizebox{\columnwidth}{!}{
    \begin{tabular}{c|cccccc}
    \hline
     &\textbf{Data} & \begin{tabular}[c]{@{}c@{}}\textbf{SNR}\\ \textbf{(dB)}\end{tabular} &\textbf{IoU}$\uparrow$ & \textbf{F1}$\uparrow$ & \textbf{HD}$\downarrow$ & \textbf{ASSD}$\downarrow$ \\ [1mm] \hline 
    \multirow{3}{*}{\rotatebox{90}{DSWE-Net}}
    & Simulation & $\infty$ & 0.732 & 0.947 & 22.82 & 2.18 \\ 
       & Simulation & 11 & 0.706 & 0.907 & 58.91 &  2.23 \\
       & CIRS049 & - &0.667 & 0.350 & 90.86 & 76.95\\
        \hline
        
    \multirow{3}{*}{\rotatebox{90}{\textbf{Ours}, $\mathrm{M}$}}
    & Simulation & $\infty$ & 0.964 & 0.981 & 1.33 & 0.130 \\ 
       & Simulation & 11 & 0.949 & 0.973 & 1.93 &  0.184 \\
       & CIRS049 & - &0.738 & 0.843 & 8.97 & 1.011\\
        \hline
    \end{tabular}}
    \label{CIRS049}
\end{table}

Table \ref{table:Masks} indicates the segmentation performance of both DSWE-Net \cite{ahmed2021dswe} and our proposed method, specifically the mask M generated from our denoiser model. The clean and noisy simulation datasets yield high IoU (0.964 and 0.949, respectively) and F1 scores (0.981 and 0.973, respectively). These metrics are quite high, making the resulting segmentation masks almost identical to the simulation phantom structures and hence, the ground truth masks. As such, the simulation masks have not been included in the corresponding visual results. We also compute the overall HD and ASSD which are more sensitive to the mask surface compared to IoU and F1-score. On average, the maximum pixel level surface gaps between the clean and simulation test cases are quite low (1.33 and 1.93, respectively). These low values make the ASSD values quite low as well (0.130 and 0.184, respectively).

For the CIRS049 phantoms, the last row in figure \ref{result_comp_cirs} includes some cases of segmentation masks. The model can identify smaller regions of inclusions in larger areas of background, regardless of their stiffness. The segmentation mask, M, is the output from a sigmoid activation function. The mask results shown in figure \ref{result_comp_cirs} are the direct output from that function without being thresholded. This means that the clean FG-BG boundaries of the masks possess high confidence. The mean IoU, F1, HD, and ASSD on the overall denoised CIRS049 test samples are respectively $0.738$, $0.843$, $8.972$, and $1.011$, as shown in table \ref{table:Masks}. The misalignment between the GT mask and estimated mask in the small inclusion cases lowers the IoU to some extent (further explained in section \ref{discussion}). 

The performance of DSWE-Net \cite{ahmed2021dswe} segmentation is inferior even in the noiseless simulation compared to our CIRS049 masks. Smears in the masks (see figure \ref{result_comp_cirs}) are responsible for this. Furthermore, the network proposed by Neidhardt et al. \cite{neidhardt2022ultrasound} neither generates segmentation masks nor provides any segmentation strategies for comparison purposes.

% However, considering Ahmed et al. \cite{ahmed2021dswe} achieved an average IoU of $0.81$ on their COMSOL-generated simulated data (only $0.072$ more), the 0.738 IoU in CIRS049 data for our method shows better potential. 

\begin{table*}[t]

\caption{Comparison of training setup: variations 1-4 are for the reconstruction network, $\mathcal{F}_{R}(\cdot)$, and variations 5-11 are for the loss coupling factors of the denoising network, $\mathcal{F}_{D}(\cdot)$, on noisy simulation test data.}
    \centering
    % \small
    \footnotesize
    % \begin{tabular}{|c|c|m{2.2cm}|m{2.2cm}|c|c|c|c|c|}
    \label{ablation}
% \resizebox{\columnwidth}{!}{
\begin{tabular}{c|lrrrrr}
\hline
\textbf{Variations} & 
        \textbf{Training Setup} &
        \textbf{MAE}$\downarrow$ & \textbf{MAE}$\downarrow$ & \textbf{{PSNR}}$\uparrow$  & \textbf{PSNR}$\uparrow$ & \textbf{PSNR}$\uparrow$ \\
         & 
         % $\alpha$: FG Loss, $\beta$: BG Loss, $\lambda$: Fusion Loss,
             &\textbf{(FG)} & \textbf{(BG)} &  & \textbf{(FG)} &  \textbf{(BG)}\\ 
         &
         % $\eta$: NCC Loss, $\gamma$: TV-Loss
         &\textbf{[kPa]} & \textbf{[kPa]} & \textbf{[dB]}  & \textbf{[dB]} &  \textbf{[dB]}\\ \hline 
1& Global Attention &  1.84  &  0.77    & 29.04 & 25.44 & 26.99  \\ [0.1mm]
2& Local Attention &  1.80  &  0.82    & 27.09 & 23.59 & 25.83  \\ [0.1mm]
3& Global $\rightarrow$ Local Attention \textbf{*}&  1.79  &  0.69    & 29.23 & 25.51 & 27.35  \\ [0.1mm]
4& Local $\rightarrow$ Global Attention &  1.88  &  0.77    & 28.00 & 24.22 & 26.95  \\ [1mm] \hline

5& $\alpha=100, \beta = 17, \lambda=0, \eta=50,  \gamma=10 $  \ &  19.38   &   10.62    & 19.92 & 14.95 & 16.03  \\ [0.1mm]
6& $\alpha=100, \beta = 17, \lambda=50, \eta=50,  \gamma=10 $  \ &  5.61   &   1.04    & 20.43 & 15.30 & 17.60  \\ [0.1mm]
7& $\alpha=0, \beta = 0, \lambda=50, \eta=50,  \gamma=10 $  \ &  6.12  & 0.99     & 19.19 & 14.86 & 16.58  \\ [0.1mm] 
8& $\alpha=100, \beta = 17, \lambda=50, \eta=0,  \gamma=10 $  \ &  6.00   &   1.03    & 20.25 & 14.73 & 16.81  \\ [0.1mm] 
9 & $\alpha=100, \beta = 17, \lambda=50, \eta=50,  \gamma=100 $\ \textbf{*} &  5.68   &   0.99    & 21.11 & 15.58 & 17.84  \\ [0.1mm]
10 & $\alpha=100, \beta = 17, \lambda=50, \eta=50,  \gamma=1000 $  \ &     6.21    & 1.04 & 21.43 & 15.98 & 19.08  \\ [0.1mm]
11& $\alpha=100, \beta = 17, \lambda=200, \eta=50,  \gamma=100 $  \ & 5.95 &  1.04 &  20.32 & 14.86 & 17.59 \\ [0.1mm] 

% 4& $\alpha=10, \beta = 10, \lambda=50, \eta=50,  \gamma=10 $  \ &  5.58   &   1.07    & 18.92 & 15.06 & 17.18  \\ [1mm] 

\hline

    \end{tabular}
    % }% 
\end{table*}

\subsection{Ablation Study}

An ablation study on the impact of the attention (global attention and local spatial grid attention) schemes and varying denoising loss coupling factors is shown in table \ref{ablation}.

The utilization of global and local attention, depicted in figure \ref{Conv-Vit}, is a significant part of our architecture.  Global attention alone achieves moderate performance (FG MAE = \(1.84\) kPa, BG MAE = \(0.77\) kPa, FG/BG PSNR = \(25.44\)/\(26.99\) dB), effectively capturing broad spatial relationships but struggling to suppress localized noise, particularly in background regions. In contrast, local attention without global context degrades reconstruction quality, increasing BG MAE to \(0.82\) kPa and reducing FG PSNR to \(23.59\) dB, as it lacks the contextual awareness necessary to distinguish noise from actual shear wave structures.

The hierarchical `Global~\(\rightarrow\)~Local' approach achieves the best performance, yielding the lowest BG MAE (\(0.69\) kPa) and highest FG/BG PSNR (\(25.51\)/\(27.35\) dB). This strategy first extracts global tissue stiffness trends, such as wavefront directionality, and subsequently refines local texture details, ensuring effective reconstruction while preserving structural integrity. Conversely, reversing this order ('{Local}~\(\rightarrow\)~Global') increases FG MAE (\(1.88\) kPa) and reduces FG PSNR (\(24.22\) dB), as prior local processing amplifies noise and degrades the global context necessary for accurate stiffness estimation. These impacts on the reconstruction performance highlight the superiority of a hierarchical \textbf{Global}~\(\rightarrow\)~\textbf{Local} strategy which was finally selected for our network which is marked with a star (\textbf{*}) in table \ref{ablation}.

The influences of the multiple loss components are investigated as well in table \ref{ablation}. When the coupling factor for the fusion loss (\(\lambda\)) is excluded (variation 5), the network performs poorly with the highest MAE values for both FG and BG, and the lowest PSNR values. This makes sense as the fused result is the main output of the denoiser network. Including \(\lambda\) (variation 6 onwards) considerably improves performance, reducing MAE and increasing PSNR values. Omitting either \{$\alpha, \beta$\} or \(\eta\) (FG-BG loss or NCC loss) degrades the MAE compared to them being included, as seen from variation 10 and 11. Setting \(\gamma\) (TV loss coupler) to a high value like 1000 (variation 10) improves PSNR slightly but increases the MAE for FG, suggesting a diminishing return when TV loss dominates. Moderate values of \(\gamma\) (e.g., variation 9) provide a balanced improvement, achieving the highest PSNR for FG and BG. From these results, variation-9 is taken as an optimum choice for the coupling values. The results indicate that balanced coupling for TV-loss leads to the best performance, while excluding key components (like \(\alpha, \beta, \lambda\)) negatively impacts denoising efficacy.

\section{Discussion}\label{discussion}

This study presents a novel transformer-based deep learning framework designed for accurate modulus estimation, denoising, and segmentation in shear wave elastography.  Our combined framework of Convolutional Neural Networks (CNNs) and Vision Transformer (ViT) architectures, leverages volumetric shear wave data to achieve robust performance. Our approach utilizes a two-stage network: (i) a reconstruction network and (ii) a denoiser network. The reconstruction network utilizes multi-resolution Spatio-Temporal ViT modules (STViT) within an encoder-decoder architecture.  This design facilitates optimal mapping of 3D motion data into a 2D modulus estimation with low noise and low variation in the stiffness map. The quality of this generated modulus was presented in sections \ref{Simu_performance} and \ref{CIRS_performance} both qualitatively and quantitatively. Section \ref{mask_performance} discusses the quantitative performance metrics of the generated segmentation masks for our approach.

% Table without Median
\begin{table*}[!t]
\centering
\footnotesize
\renewcommand{\arraystretch}{1}
\caption{Quantitative comparison (mean, std. kPa) of various techniques for \textcolor{blue}{11dB SNR simulation} test samples depicted in figure \ref{result_comp_simu}.}
\label{table:test_results_noisy}

\begin{tabular}{ccccccccccc}
\hline
\textbf{Cases} & \textbf{Method} & \textbf{FG} & \textbf{BG} & \textbf{FG Mean [kPa]} & \textbf{BG Mean [kPa]} & \textbf{PSNR}$\uparrow$ & \textbf{CNR}$\uparrow$ & \textbf{SSIM}$\uparrow$\\
 & & \textbf{[kPa]} & \textbf{[kPa]} & \textbf{$\pm$ STD$\downarrow$ [kPa]} & \textbf{$\pm$ STD$\downarrow$ [kPa]} & \textbf{[dB]} & \textbf{[dB]} & \\
\hline
\multirow{3}{*}{\textbf{A}} & Neidhardt et al. \cite{neidhardt2022ultrasound} & \multirow{3}{*}{61} & \multirow{3}{*}{14} & 66.42 $\pm$ 5.05 & 14.43 $\pm$ 0.87  & 20.09 & 28.03 & 0.970\\
& DSWENet \cite{ahmed2021dswe} & & & 61.68 $\pm$ 9.36 & 15.86 $\pm$ 4.79  &15.72 & 25.78 & 0.953\\
% & Ours & & & $61.75 \pm 2.70$ & $13.99 \pm 0.46$ & $61.54$ & $13.93$ \\
& Ours ($\mathrm{Y}$) & & & 63.44 $\pm$ 2.03 & 13.89 $\pm$ 0.24 & 40.80 & 42.02 & 0.998\\
\hline

\multirow{3}{*}{\textbf{B}} & Neidhardt et al. \cite{neidhardt2022ultrasound} & \multirow{3}{*}{89} & \multirow{3}{*}{25} & 90.30 $\pm$  4.23 & 25.43 $\pm$ 0.81 & 27.99 & 34.80 & 0.967 \\
& DSWENet \cite{ahmed2021dswe} & & & 79.49 $\pm$ 10.47 & 26.75 $\pm$ 4.01 & 22.98& 33.80 & 0.95\\
% & Ours & & & $88.80 \pm 2.31$ & $25.19 \pm 0.67$ & $89.20$ & $25.11$ \\
& Ours ($\mathrm{Y}$) & & & 90.69 $\pm$ 0.90 & 25.14 $\pm$ 0.29 & 43.21 & 47.48 & 0.999\\
\hline

\multirow{3}{*}{\textbf{C}} & Neidhardt et al. \cite{neidhardt2022ultrasound} & \multirow{3}{*}{9} & \multirow{3}{*}{20} & 9.89 $\pm$ 0.60 & 20.27 $\pm$ 0.67 & 29.62 & 38.74 & 0.958\\
& DSWENet \cite{ahmed2021dswe} & & & 11.96 $\pm$ 2.05 & 21.66 $\pm$ 1.70 & 28.71 & 33.60 & 0.953\\
% & Ours & & & $9.59 \pm 0.40$ & $20.05 \pm 0.22$ & $9.51$ & $20.04$ \\
& Ours ($\mathrm{Y}$) & & & 10.82 $\pm$ 0.38 & 20.00 $\pm$ 0.08  & 48.40 & 45.39 & 0.999\\
\hline

\multirow{3}{*}{\textbf{D}} & Neidhardt et al. \cite{neidhardt2022ultrasound} & \multirow{3}{*}{91} & \multirow{3}{*}{24} & 81.09 $\pm$ 2.83 & 24.29 $\pm$ 0.65 & 16.84 & 23.72 & 0.968\\
& DSWENet \cite{ahmed2021dswe} & & & 60.93 $\pm$ 9.02 & 25.44 $\pm$ 2.06  & 13.83 & 18.46 & 0.948\\
% & Ours & & & $85.08 \pm 2.67$ & $24.14 \pm 0.35$ & $85.60$ & $24.08$ \\
& Ours ($\mathrm{Y}$) & & & 87.35 $\pm$ 0.91 & 24.08 $\pm$ 0.09   &26.01&38.58& 0.983 \\
\hline

\multirow{3}{*}{\textbf{E}} & Neidhardt et al. \cite{neidhardt2022ultrasound} & \multirow{3}{*}{61} & \multirow{3}{*}{23} & 56.82 $\pm$ 3.26 & 23.29 $\pm$ 0.61 & 21.32 & 35.45 & 0.902\\
& DSWENet \cite{ahmed2021dswe} & & & 55.11 $\pm$ 7.09 & 25.02 $\pm$ 1.78  & 18.23 & 27.68 & 0.885\\
% & Ours & & & $62.02 \pm 1.81$ & $23.15 \pm 0.17$ & $62.50$ & $23.12$ \\
& Ours ($\mathrm{Y}$) & & & 61.98 $\pm$ 0.61 & 23.12 $\pm$ 0.06  &36.51&39.02& 0.992\\
\hline

\multirow{3}{*}{\textbf{F}} & Neidhardt et al. \cite{neidhardt2022ultrasound} & \multirow{3}{*}{9} & \multirow{3}{*}{27} & 10.56 $\pm$ 0.65 & 27.34 $\pm$ 0.66  & 25.03 & 37.98 & 0.916 \\
& DSWENet \cite{ahmed2021dswe} & & & 14.57 $\pm$ 3.39 & 28.14 $\pm$ 1.31  & 28.24 & 35.42& 0.945\\
% & Ours & & & $9.53 \pm 0.25$ & $27.01 \pm 0.17$ & $9.49$ & $27.02$ \\
& Ours ($\mathrm{Y}$) & & & 9.74 $\pm$ 0.75 &26.97  $\pm $ 0.09  &44.83&40.54& 0.997\\
\hline

\end{tabular}
\end{table*}

\begin{table*}[!t]
\centering
\footnotesize
\renewcommand{\arraystretch}{1}
\caption{Quantitative comparison (mean, std. kPa) of various techniques for \textcolor{blue}{CIRS049 phantom} test samples depicted in figure \ref{result_comp_cirs}}
\label{table:test_results_CIRS}

\begin{tabular}{ccccccccccc}
\hline
\textbf{Cases} & \textbf{Method} & \textbf{FG} & \textbf{BG} & \textbf{FG Mean [kPa]} & \textbf{BG Mean [kPa]} & \textbf{PSNR}$\uparrow$ & \textbf{CNR}$\uparrow$ & \textbf{SSIM}$\uparrow$\\
 & & \textbf{[kPa]} & \textbf{[kPa]} & \textbf{$\pm$ STD$\downarrow$ [kPa]} & \textbf{$\pm$ STD$\downarrow$ [kPa]} & \textbf{[dB]} & \textbf{[dB]} & \\
\hline
\multirow{3}{*}{\textbf{A}} & Neidhardt et al. \cite{neidhardt2022ultrasound} & \multirow{3}{*}{12} & \multirow{3}{*}{24} & 9.91 $\pm$ 1.30 & 23.45 $\pm$ 1.02 & 15.82 & 22.40 & 0.943 \\
 & DSWENet \cite{ahmed2021dswe} & & & 11.32 $\pm$ 1.45 & 23.56 $\pm$ 12.13  &18.24 & 23.58 & 0.946\\
 % & Ours & & & 9.83 $\pm$ 0.86 & 24.02 $\pm$ 0.18 & 9.79 & 24.02 \\
 % & Ours (Patched) & & & 11.70 $\pm$ 0.93 & 23.99 $\pm$ 0.35 & 12.08 & 24.04 \\
 & Ours ($\mathrm{Y}$) & & & 11.69 $\pm$ 0.64 & 24.06 $\pm$ 0.48  &22.38 &36.92 & 0.962\\
\hline

\multirow{3}{*}{\textbf{B}} & Neidhardt et al. \cite{neidhardt2022ultrasound} & \multirow{3}{*}{39} & \multirow{3}{*}{24} & 42.35 $\pm$ 15.15 & 23.96 $\pm$ 1.11  & 12.13 & 24.31 & 0.907\\
 & DSWENet \cite{ahmed2021dswe} & & & 53.65 $\pm$ 7.14 & 24.87 $\pm$ 1.77  &11.29 &29.26 & 0.913 \\
 % & Ours & & & 38.27 $\pm$ 1.75 & 23.99 $\pm$ 0.32 & 37.75 & 24.00 \\
 % & Ours (Patched) & & & 42.08 $\pm$ 7.81 & 24.04 $\pm$ 0.24 & 39.08 & 24.04 \\
 & Ours ($\mathrm{Y}$) & & & 42.28 $\pm$ 5.36 & 23.98 $\pm$ 0.06   &14.13 &45.95 & 0.962 \\
\hline

\multirow{3}{*}{\textbf{C}} & Neidhardt et al. \cite{neidhardt2022ultrasound} & \multirow{3}{*}{66} & \multirow{3}{*}{24} & 66.07 $\pm$ 3.20 & 23.20 $\pm$ 1.78  & 18.61 & 27.61 & 0.902 \\
 & DSWENet \cite{ahmed2021dswe} & & & 52.67 $\pm$ 4.56 & 23.55 $\pm$ 1.23  &19.87 &25.92 & 0.854 \\
 % & Ours & & & 55.21 $\pm$ 2.65 & 23.17 $\pm$ 0.85 & 55.21 & 23.48 \\
 % & Ours (Patched) & & & 66.50 $\pm$  1.52 &  23.74 $\pm$  0.85 & 66.54 & 24.03 \\
 & Ours ($\mathrm{Y}$) & & & 65.82 $\pm$ 1.00 & 23.96 $\pm$ 0.13   &22.93 &45.50 & 0.946 \\
\hline

\multirow{3}{*}{\textbf{D}} & Neidhardt et al. \cite{neidhardt2022ultrasound} & \multirow{3}{*}{36} & \multirow{3}{*}{21} & 44.64 $\pm$ 10.38 & 20.31 $\pm$ 0.87  & 10.09 & 25.75 & 0.864 \\
 & DSWENet \cite{ahmed2021dswe} & & & 52.44 $\pm$ 11.42 & 19.87 $\pm$ 1.38  &10.06 &29.04 & 0.778 \\
 % & Ours & & & 44.57 $\pm$ 4.41 & 20.83 $\pm$ 0.78 & 44.49 & 21.23 \\
 % & Ours (Patched) & & & 36.32 $\pm$ 1.74 & 20.76 $\pm$ 0.78 & 36.14 & 20.97 \\
 & Ours ($\mathrm{Y}$) & & & 36.45 $\pm$ 0.61 & 20.66 $\pm$ 0.07  &21.34 &41.60 & 0.942 \\
\hline

\multirow{3}{*}{\textbf{E}} & Neidhardt et al. \cite{neidhardt2022ultrasound} & \multirow{3}{*}{72} & \multirow{3}{*}{20} & 66.97 $\pm$ 5.87 & 20.32 $\pm$ 1.10 & 14.72 & 28.00 & 0.708\\
 & DSWENet \cite{ahmed2021dswe} & & & 80.53 $\pm$ 14.34 & 20.55 $\pm$ 1.75  &19.45 &29.81 & 0.846 \\
 % & Ours & & & 59.58 $\pm$ 9.64 & 19.34 $\pm$ 1.21 & 59.23 & 19.26 \\
 % & Ours (Patched) & & & 65.23 $\pm$ 0.86 &  23.11 $\pm$ 1.86 & 65.09 & 23.91 \\
 & Ours ($\mathrm{Y}$) & & & 67.21 $\pm$ 0.70 &  23.84 $\pm$ 0.42  &26.53 &47.73 & 0.957 \\
\hline

\end{tabular}
\end{table*}

\begin{figure*}[t]
  \centering
  \includegraphics[width=1\textwidth]{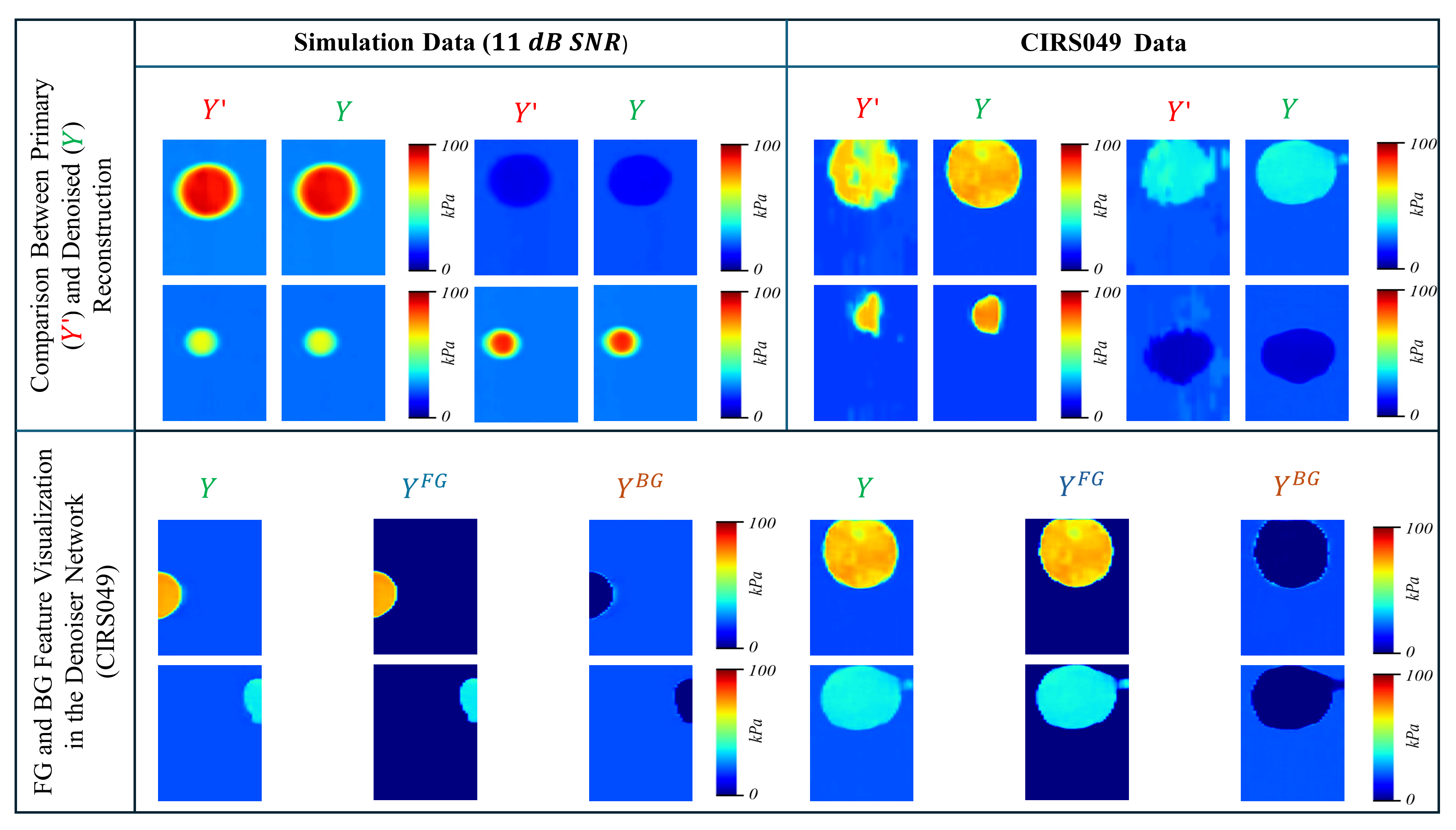}
  \caption{Comparison between the primary reconstruction $\text{Y}'$, foreground feature $Y^{FG}$ and background feature $Y^{BG}$ with respect to the corresponding cleaned output, $\text{Y}$ from our pipeline.  } 
  \label{ALL_intermediate_features} 
\end{figure*}

Stiffness estimations for the six simulated cases (A–F) under $11$~dB SNR, shown in Figure~\ref{result_comp_simu}, are summarized in Table~\ref{table:test_results_noisy}. These cases differ in phantom size, FG-BG stiffness difference, and inclusion location. Across all cases, our method outperforms Neidhardt et al.~\cite{neidhardt2022ultrasound} in most metrics, including both mean and standard deviation, which is consistent with the visual trends observed in Figure~\ref{result_comp_simu}. Neidhardt et al.'s method exhibits both overestimation (Cases A and F) and underestimation (Cases D and E) of FG stiffness, with deviations ranging from $1$ to $5$~kPa from ground truth. DSWENet~\cite{ahmed2021dswe} performs even worse under noise, showing deviations up to $\pm9$~kPa in cases like D. In contrast, our approach remains robust under noisy conditions, delivering accurate estimates with consistently low standard deviation. This robustness is largely attributed to the integration of a post-denoiser module, which further refines the initial modulus maps generated by the reconstruction network.

\begin{table}[t]

\caption{Performance metric comparison between the primary reconstructions and denoised outputs}
    \centering
    \footnotesize
    % \begin{tabular}{|c|c|m{2.2cm}|m{2.2cm}|c|c|c|c|c|}
    \label{Y_prime and Y result comparison}
    \begin{tabular}{ccrrr}
    \hline
        \textbf{Data} & \textbf{Feature} & \textbf{MAE}$\downarrow$ & \textbf{MAE}$\downarrow$ &  \textbf{PSNR}$\uparrow$ \\

         & & \textbf{(FG)[kPa]} & \textbf{(BG)[kPa]} &  \textbf{[dB]}  \\ \hline
        
        \multirow{4}{*}{}\textbf{Simulation} & $\mathrm{Y}'$ & 1.84 & 0.24 & 28.37 \\  
        (11~dB SNR) & $\mathrm{Y}$ & 1.42 & 0.16 &  32.68 \\ 
        \hline

       \multirow{4}{*}{}\textbf{CIRS049} & $\mathrm{Y}'$ & 6.09 & 1.31 & 19.75 \\ 
        % \textbf{} & $\mathbf{Y}$  & 5.34 & 1.34  & 20.83 \\ \hline
        \textbf{} & $\mathrm{Y}$  & 5.68 & 0.99  & 21.11 \\ \hline

    \end{tabular}
\end{table}

The five CIRS049 phantom cases shown in Figure~\ref{result_comp_cirs} are quantitatively summarized in Table~\ref{table:test_results_CIRS}. Our method consistently outperforms Neidhardt et al.~\cite{neidhardt2022ultrasound} and DSWENet~\cite{ahmed2021dswe} across five different cases. Similar to the simulation results, our approach achieves notably lower standard deviations, indicating more consistent stiffness estimation. For example, in Case A, our method yields FG and BG mean values of $11.69$~kPa and $24.06$~kPa, respectively, with low variability ($\pm0.64$~kPa and $\pm0.48$~kPa), closely matching the reference values. In Case C, while Neidhardt et al. slightly improve the FG mean ($66.07$~kPa) over our estimate ($65.82$~kPa), their result suffers from a higher standard deviation and lower PSNR ($18.61$~dB vs. $22.93$~dB), suggesting noisier predictions. In Case E, our SSIM of $0.957$ indicates superior structural preservation compared to other methods. Both Neidhardt et al.~\cite{neidhardt2022ultrasound} and DSWENet~\cite{ahmed2021dswe} exhibit higher standard deviations across key performance metrics.

% The denoising is guided by a multi-task loss function defined in equation (\ref{denoise}). 

Figure \ref{ALL_intermediate_features} presents a comparison between the primary reconstructions ($\mathrm{Y'}$) and the denoised final outputs ($\mathrm{Y}$). It also illustrates the individually refined FG and BG ($Y^{FG}$ and $Y^{BG}$) on both the simulation (SNR $11$~dB) and CIRS049 dataset. The $\mathrm{Y'}$ demonstrates accurate mean estimations but a higher standard deviation compared to the respective $\mathrm{Y}$. The denoiser network is trained to achieve two main objectives: (i) fixing any shape distortions and noise prevalence, and (ii) accurately retaining the modulus distribution of $\mathrm{Y}$. Consequently, any high deviations in the $\mathrm{Y'}$ estimations from the reconstruction network can lead to under- or over-estimated clean outputs ($\mathrm{Y}$). Nevertheless, the denoiser network effectively compensates for abrupt variations in the FG and BG regions, considerably enhancing the primary reconstruction. The overall improvement in reconstruction achieved by the denoiser network is detailed in table \ref{Y_prime and Y result comparison}.

The foreground $Y^{FG}$ and background $Y^{BG}$ were directly supervised using the loss terms $\mathcal{L}_{FG}$ and $\mathcal{L}_{BG}$ (see equations (\ref{FG_loss}) and (\ref{BG_loss})), respectively. To investigate whether the corresponding losses have performed in an anticipated manner, we present some test cases from the CIRS049 dataset in figure \ref{ALL_intermediate_features} showing their corresponding $\text{Y}$, $Y^{FG}$, and $Y^{BG}$. We see that $Y^{FG}$ contains only the foreground estimations with the background values zeroed out. Similarly, in the $Y^{BG}$, only the background values exist, with foreground values at zero. The fusion blocks take the denoised features, $F_2$ and $B_2$, as inputs to generate a clean output estimation, $\text{Y}$ and the segmentation mask, $\text{M}$. As can be seen from figure \ref{ALL_intermediate_features}, the fusion was performed very decently.

\begin{figure*}[t]
  \centering
  \includegraphics[width=0.7\textwidth]{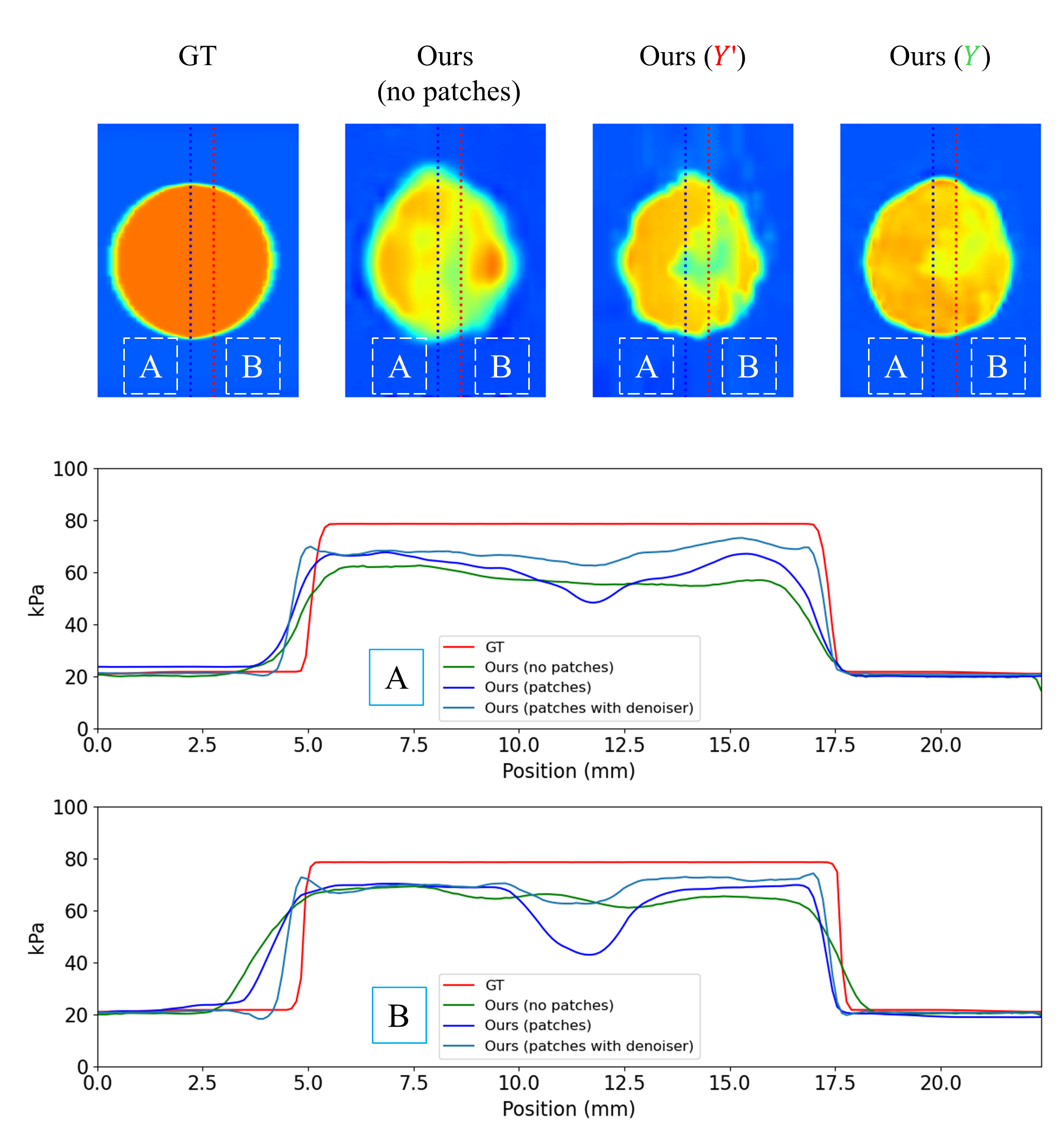}
  \caption{Comparison of patch-based training in limited CIRS049 phantom data using two axial slices (A and B).} 
  \label{patch_no_patch} 
\end{figure*}

% has a larger inclusion area than the ground truth across all models. 

% Due to the notable stiffness difference in CIRS049 Case E (FG: 72 kPa, BG: 20 kPa), the MAE is high when the ground truth BG around the inclusion overlaps with the estimated FG. Such outliers have inflated the overall MAE of CIRS049 test cases.

\begin{table}[t]
\centering
\setlength{\tabcolsep}{4pt} % Reduce column spacing
% \scriptsize % Reduce font size
\footnotesize
\caption{Comparison between patched and non-patched for our method with CIRS dataset [$\uparrow$: higher is better, $\downarrow$: lower is better]}
\label{patch result comparison}
\begin{tabular}{ccrrr}
    \hline
    \textbf{Data} & \textbf{Method} & \textbf{MAE}$\downarrow$ & \textbf{MAE}$\downarrow$  & \textbf{PSNR}$\uparrow$ \\
     &&  \textbf{(FG)} & \textbf{(BG)} &    \\
     &&   \textbf{[kPa]} & \textbf{[kPa]}  & \textbf{[dB]}   \\
    \hline
    \multirow{2}{*}{\textbf{CIRS049}}  & Non-Patched & 8.30 & 1.55  & 15.73  \\
    & Patched & 5.68 & 0.99  & 21.11  \\
    \hline
\end{tabular}
\end{table}

It is crucial to highlight that the denoiser network's function is to refine the 2D phantom image from the primary reconstruction network. It does not have information about 3D motion propagation and related features and as such, cannot correct drastic shape changes or notable deviations in reconstructions. If the mean estimate deviates from the ground truth, the refined output, while visually cleaner, will also be inaccurate. To achieve accurate mapping of $\text{Y}'$ to the ground label, proper training of the primary reconstruction network is essential. Additionally, the refinement of the denoiser is based on its training data. For instance, a denoiser trained on irregular shapes will produce different results from one trained on circular phantoms, which tends to make outputs more circular (see figure \ref{patch_no_patch}). It is worth mentioning that the primary reconstruction $\text{Y}'$ alone can surpass those of previously reported deep learning techniques in realistic scenarios. For instance, in noisy simulations, the $\text{Y}'$ foreground MAE is reduced to 1.84 kPa, with a background MAE of 0.24 kPa. Similarly, experiments with CIRS049 phantoms $\text{Y}'$ demonstrate notable improvements, with a foreground MAE of 6.09 kPa and a background MAE of 1.31 kPa (see table \ref{Y_prime and Y result comparison}).

The relatively higher MAE of inclusions in the CIRS049 phantom data (see table \ref{Test cases result comparison}) stems from minor inconsistencies in the CIRS049 data's ground truth labels. These labels were manually generated using CVAT software on B-mode images, with boundaries placed where transitions between FG and BG regions were visually apparent. However, some B-mode images lacked distinct transitions, resulting in labels inconsistent with the actual modulus mapping, leading to overlaps between the actual FG and BG. Given the higher stiffness in FG regions, such overlaps produce higher MAE values. This inconsistency is evident in cases where all the models (including ours) generated comparable phantom borders that differed from the ground truth. For example, in figure \ref{result_comp_cirs} (Case: D), the reconstructed YM maps from each technique are shifted lower than the ground truth.

An important point to be mentioned is that the reconstruction effectiveness on noise-added simulation data, as well as CIRS049 tissue mimicking experimental data, does not fully account for the practical challenges during $in-vivo$ shear wave elastography. Clinical SWE data, which are otherwise unavailable to us, can have various irregularity and heterogeneity (from tissue, muscle, lesions, mass, tumor, etc.) compared to experimental data. And, the noisy conditions will be vastly different from the ones we have investigated in this study, affecting the training process. In the future, we aim to evaluate as well as tweak our method with clinical SWE data. This will largely improve the generalization capability of our method with the potential for real-world deployment.

In multi-push-based SWE data, the ROIs do not need to be as large as those required for single-push data. This is because the decayed intensity of one ARF can be compensated by the others, resulting in a notable advantage for multi-push data. Specifically, multi-push data allows for an increase in the number of ROIs, which in turn leads to more training samples. Furthermore, employing a patched configuration during training can considerably boost the amount of training data, which is especially beneficial when dealing with limited datasets—a common challenge in real-world clinical scenarios. This issue of limited samples was encountered during the CIRS049 phantom training, where the use of patch-based training greatly improved the reconstruction quality, as demonstrated in figure \ref{patch_no_patch}. Two axial slices, labeled \textit{A} and \textit{B}, were analyzed to compare the ground truth with the estimated reconstructions. While training on multi-push data alone produced smooth estimates, the inclusion values showed notable errors. The average MAE for the FG and BG from multi-push data without patches were $8.30$~kPa and $1.55$~kPa, respectively. In contrast, sequential multi-push training with patches reduced the FG MAE to $5.68$~kPa and the BG MAE to $0.99$~kPa (see table \ref{patch result comparison}).

Moreover, an examination of slice \textit{B} from figure \ref{patch_no_patch}, which highlights a thinner part of the inclusion, reveals that patch-based training more closely follows the ground truth compared to the non-patched version. However, an underestimation is observed in the center of both slices A and B. This issue arises due to destructive interference caused by reflection at the center, where smaller patches fail to compensate for the lack of information—a challenge that non-patch-based training can overcome due to its larger spatial receptive field. Nevertheless, the inclusion of a denoiser can effectively mitigate most of these shortcomings.

\section{Conclusion}\label{discussion}
In this work, we have proposed SW-Vit pipepline which is a 3D CNN encoded Spatio-Temporal Conv-ViT network followed by a noise-resilient denoiser model for producing minimal error and high-quality SWE reconstructions as well as segmentation from simulated and experimental phantom data. The CNN encoded Spatio-Temporal Conv-ViT acts as the primary reconstruction model to map 3D motion data to 2D modulus images. The subsequent denoiser model is supervised with a compound loss consisting of region-based mapping terms, fused region supervision, TV loss term, and an IoU loss term. Such supervision enables the denoiser to take in the primary reconstruction as input to produce not only cleaner images but also corresponding segmentation masks. A patch-based training method is also introduced to retain robustness in case of scarce SWE data samples. Our method has been tested on COMSOL simulated as well as CIRS049 phantoms with different shapes, stiffness, and locations. The resulting estimations have been found to be superior to the ones generated from previously reported deep learning SWE estimation techniques: DSWE-Net \cite{ahmed2021dswe} and a spatio-temporal CNN \cite{neidhardt2022ultrasound}. The simultaneously produced segmentation masks are also precise in isolating foreground and background regions, regardless of their stiffness. Such performance of our method shows promise and paves the way for investigating reconstruction as well as segmentation capabilities in $in-vivo$ SWE data (i.e., breast, liver, etc.) as a prospective future work.

\bibliographystyle{IEEEtran}
% \bibliography{IEEEabrv,Bibliography}


% \bibliographystyle{elasarticle-harv}

% % Loading bibliography database
\bibliography{main.bib}

\begin{thebibliography}{10}
\providecommand{\url}[1]{#1}
\csname url@samestyle\endcsname
\providecommand{\newblock}{\relax}
\providecommand{\bibinfo}[2]{#2}
\providecommand{\BIBentrySTDinterwordspacing}{\spaceskip=0pt\relax}
\providecommand{\BIBentryALTinterwordstretchfactor}{4}
\providecommand{\BIBentryALTinterwordspacing}{\spaceskip=\fontdimen2\font plus
\BIBentryALTinterwordstretchfactor\fontdimen3\font minus \fontdimen4\font\relax}
\providecommand{\BIBforeignlanguage}[2]{{%
\expandafter\ifx\csname l@#1\endcsname\relax
\typeout{** WARNING: IEEEtran.bst: No hyphenation pattern has been}%
\typeout{** loaded for the language `#1'. Using the pattern for}%
\typeout{** the default language instead.}%
\else
\language=\csname l@#1\endcsname
\fi
#2}}
\providecommand{\BIBdecl}{\relax}
\BIBdecl

\bibitem{handorf2015tissue}
A.~M. Handorf, Y.~Zhou, M.~A. Halanski, and W.-J. Li, ``Tissue stiffness dictates development, homeostasis, and disease progression,'' \emph{Tissue Eng Part B Rev}, vol.~21, no.~3, pp. 175--192, 2015.

\bibitem{martinez2021causal}
L.~Martinez-Vidal, V.~Murdica, C.~Venegoni, F.~Pederzoli, M.~Bandini, A.~Necchi, A.~Salonia, and M.~Alfano, ``Causal contributors to tissue stiffness and clinical relevance in urology,'' \emph{Commun Biol}, vol.~4, no.~1, p. 1011, 2021.

\bibitem{piscaglia2016ultrasound}
F.~Piscaglia, V.~Salvatore, L.~Mulazzani, V.~Cantisani, and C.~Schiavone, ``Ultrasound shear wave elastography for liver disease. a critical appraisal of the many actors on the stage,'' \emph{Ultraschall in der Medizin-European Journal of Ultrasound}, vol.~37, no.~01, pp. 1--5, 2016.

\bibitem{bob2017ultrasound}
F.~Bob, I.~Grosu, I.~Sporea, S.~Bota, A.~Popescu, A.~Sima, R.~{\c{S}}irli, L.~Petrica, R.~Timar, and A.~Schiller, ``Ultrasound-based shear wave elastography in the assessment of patients with diabetic kidney disease,'' \emph{Ultrasound Med Biol}, vol.~43, no.~10, pp. 2159--2166, 2017.

\bibitem{farrow2020novel}
M.~Farrow, J.~Biglands, A.~M. Alfuraih, R.~J. Wakefield, and A.~L. Tan, ``Novel muscle imaging in inflammatory rheumatic diseases—a focus on ultrasound shear wave elastography and quantitative mri,'' \emph{Frontiers in Medicine}, vol.~7, p. 434, 2020.

\bibitem{ferraioli2014shear}
G.~Ferraioli, P.~Parekh, A.~B. Levitov, and C.~Filice, ``Shear wave elastography for evaluation of liver fibrosis,'' \emph{J Ultrasound Med}, vol.~33, no.~2, pp. 197--203, 2014.

\bibitem{dirrichs2016shear}
T.~Dirrichs, V.~Quack, M.~Gatz, M.~Tingart, C.~K. Kuhl, and S.~Schrading, ``Shear wave elastography (swe) for the evaluation of patients with tendinopathies,'' \emph{Academic radiology}, vol.~23, no.~10, pp. 1204--1213, 2016.

\bibitem{greenleaf2003selected}
J.~F. Greenleaf, M.~Fatemi, and M.~Insana, ``Selected methods for imaging elastic properties of biological tissues,'' \emph{Annu Rev Biomed Eng}, vol.~5, no.~1, pp. 57--78, 2003.

\bibitem{shiina2015wfumb}
T.~Shiina, K.~R. Nightingale, M.~L. Palmeri, T.~J. Hall, J.~C. Bamber, R.~G. Barr, L.~Castera, B.~I. Choi, Y.-H. Chou, D.~Cosgrove \emph{et~al.}, ``Wfumb guidelines and recommendations for clinical use of ultrasound elastography: Part 1: basic principles and terminology,'' \emph{Ultrasound Med Biol}, vol.~41, no.~5, pp. 1126--1147, 2015.

\bibitem{barr2015wfumb}
R.~G. Barr, K.~Nakashima, D.~Amy, D.~Cosgrove, A.~Farrokh, F.~Schafer, J.~C. Bamber, L.~Castera, B.~I. Choi, Y.-H. Chou \emph{et~al.}, ``Wfumb guidelines and recommendations for clinical use of ultrasound elastography: Part 2: breast,'' \emph{Ultrasound Med Biol}, vol.~41, no.~5, pp. 1148--1160, 2015.

\bibitem{yoon2011interobserver}
J.~H. Yoon, M.~H. Kim, E.-K. Kim, H.~J. Moon, J.~Y. Kwak, and M.~J. Kim, ``Interobserver variability of ultrasound elastography: how it affects the diagnosis of breast lesions,'' \emph{AJR Am J Roentgenol}, vol. 196, no.~3, pp. 730--736, 2011.

\bibitem{Youk}
J.~H. Youk, H.~Gweon, and E.~Son, ``Shear-wave elastography in breast ultrasonography: state of the art,'' \emph{Ultrasonography}, vol.~36, 04 2017.

\bibitem{Woo}
S.~Woo, C.~H. Suh, S.~Y. Kim, J.~Cho, and S.~H. Kim, ``Shear-wave elastography for detection of prostate cancer: A systematic review and diagnostic meta-analysis,'' \emph{AJR Am J Roentgenol}, vol. 209, pp. 1--9, 08 2017.

\bibitem{sarvazyan1998shear}
A.~P. Sarvazyan, O.~V. Rudenko, S.~D. Swanson, J.~B. Fowlkes, and S.~Y. Emelianov, ``Shear wave elasticity imaging: a new ultrasonic technology of medical diagnostics,'' \emph{Ultrasound Med Biol}, vol.~24, no.~9, pp. 1419--1435, 1998.

\bibitem{lai2009introduction}
W.~M. Lai, D.~Rubin, and E.~Krempl, \emph{Introduction to continuum mechanics}.\hskip 1em plus 0.5em minus 0.4em\relax Butterworth-Heinemann, 2009.

\bibitem{tanter2008quantitative}
M.~Tanter, J.~Bercoff, A.~Athanasiou, T.~Deffieux, J.-L. Gennisson, G.~Montaldo, M.~Muller, A.~Tardivon, and M.~Fink, ``Quantitative assessment of breast lesion viscoelasticity: initial clinical results using supersonic shear imaging,'' \emph{Ultrasound Med Biol}, vol.~34, no.~9, pp. 1373--1386, 2008.

\bibitem{song2012comb}
P.~Song, H.~Zhao, A.~Manduca, M.~W. Urban, J.~F. Greenleaf, and S.~Chen, ``Comb-push ultrasound shear elastography (cuse): a novel method for two-dimensional shear elasticity imaging of soft tissues,'' \emph{IEEE Trans Med Imaging}, vol.~31, no.~9, pp. 1821--1832, 2012.

\bibitem{bercoff2004supersonic}
J.~Bercoff, M.~Tanter, and M.~Fink, ``Supersonic shear imaging: a new technique for soft tissue elasticity mapping,'' \emph{IEEE Trans Ultrason Ferroelectr Freq Control}, vol.~51, no.~4, pp. 396--409, 2004.

\bibitem{song2014fast}
P.~Song, A.~Manduca, H.~Zhao, M.~W. Urban, J.~F. Greenleaf, and S.~Chen, ``Fast shear compounding using robust 2-d shear wave speed calculation and multi-directional filtering,'' \emph{Ultrasound Med Biol}, vol.~40, no.~6, pp. 1343--1355, 2014.

\bibitem{palmeri2008quantifying}
M.~L. Palmeri, M.~H. Wang, J.~J. Dahl, K.~D. Frinkley, and K.~R. Nightingale, ``Quantifying hepatic shear modulus in vivo using acoustic radiation force,'' \emph{Ultrasound Med Biol}, vol.~34, no.~4, pp. 546--558, 2008.

\bibitem{mclaughlin2006shear}
J.~McLaughlin and D.~Renzi, ``Shear wave speed recovery in transient elastography and supersonic imaging using propagating fronts,'' \emph{Inverse Probl}, vol.~22, no.~2, p. 681, 2006.

\bibitem{6264136}
N.~C. Rouze, M.~H. Wang, M.~L. Palmeri, and K.~R. Nightingale, ``Parameters affecting the resolution and accuracy of 2-d quantitative shear wave images,'' \emph{IEEE Trans Ultrason Ferroelectr Freq Control}, vol.~59, no.~8, pp. 1729--1740, 2012.

\bibitem{bernal2011material}
M.~Bernal, I.~Nenadic, M.~W. Urban, and J.~F. Greenleaf, ``Material property estimation for tubes and arteries using ultrasound radiation force and analysis of propagating modes,'' \emph{J Acoust Soc Am}, vol. 129, no.~3, pp. 1344--1354, 2011.

\bibitem{chen2009shearwave}
S.~Chen, M.~W. Urban, C.~Pislaru, R.~Kinnick, Y.~Zheng, A.~Yao, and J.~F. Greenleaf, ``Shearwave dispersion ultrasound vibrometry (sduv) for measuring tissue elasticity and viscosity,'' \emph{IEEE Trans Ultrason Ferroelectr Freq Control}, vol.~56, no.~1, pp. 55--62, 2009.

\bibitem{kijanka2018robust}
P.~Kijanka, B.~Qiang, P.~Song, C.~A. Carrascal, S.~Chen, and M.~W. Urban, ``Robust phase velocity dispersion estimation of viscoelastic materials used for medical applications based on the multiple signal classification method,'' \emph{IEEE Trans Ultrason Ferroelectr Freq Control}, vol.~65, no.~3, pp. 423--439, 2018.

\bibitem{brum2014vivo}
J.~Brum, M.~Bernal, J.-L. Gennisson, and M.~Tanter, ``In vivo evaluation of the elastic anisotropy of the human achilles tendon using shear wave dispersion analysis,'' \emph{Phys Med Biol}, vol.~59, no.~3, p. 505, 2014.

\bibitem{kijanka2018local}
P.~Kijanka and M.~W. Urban, ``Local phase velocity based imaging: A new technique used for ultrasound shear wave elastography,'' \emph{IEEE Trans Med Imaging}, vol.~38, no.~4, pp. 894--908, 2018.

\bibitem{kijanka2019fastt}
------, ``Fast local phase velocity-based imaging: Shear wave particle velocity and displacement motion study,'' \emph{IEEE Trans Ultrason Ferroelectr Freq Control}, vol.~67, no.~3, pp. 526--537, 2019.

\bibitem{gao2020robust}
F.~Gao, X.~Liu, and Z.~Yan, ``Robust strain elastography imaging using deep learning-based denoising,'' \emph{IEEE Trans Biomed Eng}, vol.~67, no.~5, pp. 1354--1362, 2020.

\bibitem{liu2019deep}
J.~Liu, H.~Zhang, and Z.~Li, ``Deep learning-based noise reduction in ultrasound strain elastography,'' \emph{Med Image Anal}, vol.~58, p. 101541, 2019.

\bibitem{ma2021deep}
J.~Ma, K.~Zhang, and S.~Wang, ``Deep learning for improved noise robustness in ultrasound elastography,'' \emph{Ultrasound Med Biol}, vol.~47, no.~4, pp. 1023--1032, 2021.

\bibitem{jush2022deep}
F.~K. Jush, M.~Biele, P.~M. Dueppenbecker, and A.~Maier, ``Deep learning for ultrasound speed-of-sound reconstruction: Impacts of training data diversity on stability and robustness,'' \emph{arXiv preprint arXiv:2202.01208}, 2022.

\bibitem{wu2018direct}
S.~Wu, Z.~Gao, Z.~Liu, J.~Luo, H.~Zhang, and S.~Li, ``Direct reconstruction of ultrasound elastography using an end-to-end deep neural network,'' in \emph{Med Image Comput Comput Assist Interv}.\hskip 1em plus 0.5em minus 0.4em\relax Springer, 2018, pp. 374--382.

\bibitem{tripura2023wavelet}
T.~Tripura, A.~Awasthi, S.~Roy, and S.~Chakraborty, ``A wavelet neural operator based elastography for localization and quantification of tumors,'' \emph{Comput Methods Programs Biomed}, vol. 232, p. 107436, 2023.

\bibitem{tehrani2022lateral}
A.~K. Tehrani, M.~Ashikuzzaman, and H.~Rivaz, ``Lateral strain imaging using self-supervised and physically inspired constraints in unsupervised regularized elastography,'' \emph{IEEE Trans Med Imaging}, 2022.

\bibitem{delaunay2021unsupervised}
R.~Delaunay, Y.~Hu, and T.~Vercauteren, ``An unsupervised learning approach to ultrasound strain elastography with spatio-temporal consistency,'' \emph{Phys Med Biol}, vol.~66, no.~17, p. 175031, 2021.

\bibitem{ahmed2021dswe}
S.~Ahmed, U.~Kamal, and M.~K. Hasan, ``Dswe-net: A deep learning approach for shear wave elastography and lesion segmentation using single push acoustic radiation force,'' \emph{Ultrasonics}, vol. 110, p. 106283, 2021.

\bibitem{neidhardt2022ultrasound}
M.~Neidhardt, M.~Bengs, S.~Latus, S.~Gerlach, C.~J. Cyron, J.~Sprenger, and A.~Schlaefer, ``Ultrasound shear wave elasticity imaging with spatio-temporal deep learning,'' \emph{IEEE Trans Biomed Eng}, 2022.

\bibitem{inoue2017development}
M.~Inoue, K.~Itoh, R.~Nishio, M.~Yoshikawa, K.~Hayashi, H.~Yamada, K.-i. Yano, H.~Sugimoto, and T.~Nishiura, ``Development of sequential multi-push shear wave elastography for in vivo human liver tissue characterization,'' \emph{Jpn J Appl Phys}, vol.~56, no. 7S1, p. 07JF07, 2017.

\bibitem{inproceedings}
J.~Hu, L.~Shen, and G.~Sun, ``Squeeze-and-excitation networks,'' 06 2018, pp. 7132--7141.

\bibitem{zamir2022restormer}
S.~W. Zamir, A.~Arora, S.~Khan, M.~Hayat, F.~S. Khan, and M.-H. Yang, ``Restormer: Efficient transformer for high-resolution image restoration,'' in \emph{Proc IEEE/CVF Conf Comput Vis Pattern Recognit}, 2022, pp. 14\,468--14\,478.

\bibitem{nowicki2020safety}
\BIBentryALTinterwordspacing
A.~Nowicki, ``Safety of ultrasonic examinations; thermal and mechanical indices,'' \emph{Med Ultrason}, vol.~22, no.~2, p. 2372, 2020. [Online]. Available: \url{https://www.medultrason.ro/medultrason/index.php/medultrason/article/view/2372}
\BIBentrySTDinterwordspacing

\bibitem{hara2017learning}
K.~Hara, H.~Kataoka, and Y.~Satoh, ``Learning spatio-temporal features with 3d residual networks for action recognition,'' in \emph{Proc IEEE Int Conf Comput Vis Workshops}, 2017, pp. 3154--3160.

\bibitem{dosovitskiy2021image}
A.~Dosovitskiy, L.~Beyer, A.~Kolesnikov, D.~Weissenborn, X.~Zhai, T.~Unterthiner, M.~Dehghani, M.~Minderer, G.~Heigold, S.~Gelly, J.~Uszkoreit, and N.~Houlsby, ``An image is worth 16x16 words: Transformers for image recognition at scale,'' \emph{arXiv preprint arXiv:2010.11929}, 2021.

\bibitem{huttenlocher1993comparing}
D.~P. Huttenlocher, G.~A. Klanderman, and W.~J. Rucklidge, ``Comparing images using the hausdorff distance,'' \emph{IEEE Trans Pattern Anal Mach Intell}, vol.~15, no.~9, pp. 850--863, 1993.

\bibitem{heimann2009comparison}
T.~Heimann, B.~Van~Ginneken, M.~A. Styner, Y.~Arzhaeva, V.~Aurich, C.~Bauer, A.~Beck, C.~Becker, R.~Beichel, G.~Bekes \emph{et~al.}, ``Comparison and evaluation of methods for liver segmentation from ct datasets,'' \emph{IEEE Trans Med Imaging}, vol.~28, no.~8, pp. 1251--1265, 2009.

\end{thebibliography}

% % Here goes the biography details.

% % Here are the biography details.

\end{document}